\documentclass[letterpaper, 10 pt, conference]{ieeeconf}  

\IEEEoverridecommandlockouts                              
\usepackage[utf8]{inputenc}
\usepackage[T1]{fontenc}
\usepackage{amssymb}
\usepackage{amsmath}
\usepackage{amsfonts}
\usepackage{graphicx}
\usepackage{subfigure}
\usepackage{bm}

\title{\fontsize{22}{13}\selectfont
Orthogonal Constant-Amplitude Sequence
Families for System Parameter Identification in Spectrally Compact OFDM
}

\author{Shih-Hao Lu, Char-Dir Chung, Wei-Chang Chen, and Ping-Feng Tsou
\thanks{Shih-Hao Lu is with the Graduate Institute of Communication Engineering, National Taiwan University, Taipei, Taiwan 10617. E-mail: michael10330146@gmail.com.}%
\thanks{Char-Dir Chung is with the Department of Electrical Engineering, National Taiwan University, Taipei, Taiwan 10617. Telephone and Fax: +886-2-3366-3596, E-mail: cdchung@ntu.edu.tw.}%
\thanks{Wei-Chang Chen is with the Department of Electronic Engineering, National Taipei University of Technology, Taipei, Taiwan. E-mail: arthurchen0601@gmail.com.}%
\thanks{Ping-Feng Tsou is with Realtek Inc., Hsinchu, Taiwan 30076. E-mail: kevinwx12392@gmail.com.}%
}

\begin{document}

\maketitle
\thispagestyle{empty}
\pagestyle{empty}

\begin{abstract}

In rectangularly-pulsed orthogonal frequency division multiplexing (OFDM)
systems, constant-amplitude (CA) sequences are desirable to construct
preamble/pilot waveforms to facilitate system parameter identification
(SPI). Orthogonal CA sequences are generally preferred in various SPI
applications like random-access channel identification. However, the number
of conventional orthogonal CA sequences (e.g., Zadoff-Chu sequences) that
can be adopted in cellular communication without causing sequence
identification ambiguity is insufficient. Such insufficiency causes heavy
performance degradation for SPI requiring a large number of identification
sequences. Moreover, rectangularly-pulsed OFDM preamble/pilot waveforms
carrying conventional CA sequences suffer from large power spectral
sidelobes and thus exhibit low spectral compactness. This paper is thus
motivated to develop several order-$I$ CA sequence families which contain
more orthogonal CA sequences while endowing the corresponding OFDM
preamble/pilot waveforms with fast-decaying spectral sidelobes. Since more
orthogonal sequences are provided, the developed order-$I$ CA sequence
families can enhance the performance characteristics in SPI requiring a
large number of identification sequences over multipath channels exhibiting
short-delay channel profiles, while composing spectrally compact OFDM
preamble/pilot waveforms.

\vspace{-0.5\baselineskip}%

\end{abstract}

\section{Introduction}

Rectangularly-pulsed orthogonal frequency division multiplexing (OFDM)
waveforms are commonly adopted in modern wireless communication systems \cite%
{LTE}-\cite{Wifi} due to their feasibility of efficient implementation by
fast Fourier transform, easy incorporation of cyclic prefix (CP) to
facilitate initial synchronization and channel estimation, and robustness
against frequency-selective channel dispersion. In rectangularly-pulsed OFDM
systems, constant amplitude (CA) sequences are often used as the training
sequence in frequency domain to modulate uniformly spaced subcarriers and
thereby enable robust fine initial time/frequency synchronization \cite{Sync
1}-\cite{CA1} and accurate channel estimation \cite{CA1}-\cite{YL sequence}
at the receiver combating frequency-selective channel dispersion. When exact
or near orthogonality is sustained among sequences, multiple CA training
sequences are also adopted to facilitate the identification of different
system parameters for establishing initial connection, including the
identification of cell/sector/antenna, random access (RA) channel, duplex
mode, guard ratio, etc. \cite{LTE}-\cite{Wifi}, \cite{PUCCH}-\cite{RA 1}.
Two typical applications based on system parameter identification (SPI) are
RA channel identification \cite{RA 1}-\cite{RA 5} and multiple-input
multiple-output (MIMO) channel sounding \cite{YL sequence}, \cite{MIMO CS 1}-%
\cite{MIMO CS 3 (LS)}. Specifically, the received OFDM waveforms carrying
different CA sequences in frequency domain are identified by
cross-correlating \cite{RA 1}-\cite{RA 5} and despreading \cite{YL sequence}%
, \cite{MIMO CS 1}-\cite{MIMO CS 3 (LS)} the received frequency-domain
samples with all possible identification sequences, thereby enabling RA
channel identification \cite{RA 1}-\cite{RA 5} and simultaneous channel
estimation for multiple MIMO channels \cite{YL sequence}, \cite{MIMO CS 1}-%
\cite{MIMO CS 3 (LS)}, respectively. In such applications, multiple
orthogonal CA sequences are generally preferred since better sequence
identification can be achieved to ensure less false identification in RA
channel identification and mitigate the effect of pilot contamination in
simultaneous MIMO channel estimation.

In practice, Zadoff-Chu (ZC) sequences \cite{Chu}-\cite{Popovic} are
commonly used as such training/identification/ sounding sequences due to
their features of CA and zero periodic autocorrelation (ZAC) in both time
and frequency domains \cite{preamble}, \cite{CAZAC}. Particularly,
cyclically-shiftable ZC sequences are popular in SPI applications due to the
ZAC-enabled feasibility by generating all orthogonal ZC sequences through
cyclically shifting the inverse discrete-Fourier-transform (DFT) of a
single-root ZC sequence with a cyclic shifting distance (CSD) $\varpi _{%
\text{ZC}}$. However, adjacent cyclically-shiftable ZC sequences with a
small $\varpi _{\text{ZC}}$ can not be unambiguously identified by the
receiver in the presence of multiple received path signals and the timing
uncertainty under which the start time of the received useful leading path
signal is practically synchronized only within the front portion of a CP
subinterval \cite{RA 2}, \cite{PRACH Cell}. Due to such sequence
identification ambiguity \cite{PRACH Cell}, not every cyclically-shiftable
single-root ZC sequence can be adopted for SPI in the uplink cellular
environment since a minimum CSD\ $\varpi _{\text{min}}$ is required to
differentiate distinct received cyclically-shiftable sequences sent from
uplink transmitters in different locations \cite{LTE}-\cite{Wifi}, \cite{RA
1}-\cite{RA 5}, \cite{PRACH Cell}. As the cell radius is increased, a larger 
$\varpi _{\text{min}}$ is required to avoid such sequence identification
ambiguity \cite{PRACH Cell}. The latter issue results in the shortage of
adoptable orthogonal ZC sequences in many standard preamble/pilot signaling
formats for SPI \cite{LTE}-\cite{Wifi}. For example, a total of $64$ ZC
sequences are required for RA channel identification in uplink 5G-NR \cite[%
Section 6.3.3.1]{5G}, \cite{RA 1}-\cite{RA 5}. Among the various adopted
pairs of sequence length and minimum CSD, the numbers of adoptable
orthogonal ZC sequences are upper bounded by the ratio of sequence length to
minimum CSD and turn out be much smaller than $64$. Since fewer orthogonal
ZC sequences are available, RA channel identification suffers from larger
false-identifiction error (FIE) in multipath environments exhibiting
longer-delay channel profiles, thus entailing worse false identification 
\cite{RA 1}-\cite{RA 2}. As another example in 5G-NR \cite[Section 6.4.1.4.1]%
{5G}, a MIMO system is designed to receive uplink pilot waveforms from at
most $12$ transmit antennas concurrently, and thus requires up to $12$
cyclically-shiftable (single-root) ZC sequences with $\varpi _{\text{ZC}%
}\geq \varpi _{\text{min}}$ to identify and separate different uplink
channels in order to achieve high estimation accuracy in simultaneous
channel estimation (SCE) \cite{YL sequence}, \cite{MIMO CS 1}, \cite{MIMO CS
3 (LS)}. Under this setup, $\varpi _{\text{min}}$ is specified by $N/12$ for
the adopted cyclically-shiftable ZC sequences of different sequence lengths $%
N$ \cite[Section 6.4.1.4.3]{5G}. Since at most $12$ cyclically-shiftable ZC
sequences are available for all adopted pairs of sequence length and minimum
CSD \cite[Section 6.4.1.4]{5G}, cyclically-shiftable ZC sequences generated
from different (relatively prime) roots are adopted in neighboring sectors
or cells in practical cellular environments. Unfortunately, ZC sequences
generated form different root indices are nonorthogonal and entail heavy
inter-pilot interference to SCE in cellular environments \cite{MIMO CS 1}.
This causes the pilot contamination problem \cite{YL sequence}, \cite{MIMO
CS 1}-\cite{MIMO CS 3 (LS)}. To alleviate the effect of pilot contamination
in the multiple cells/sectors environment, Yu-Lee (YL) sequences are
constructed in \cite{YL sequence} from phase-rotating cyclically-shiftable
ZC sequences generated from a single root index appropriately, and shown to
outperform multiple-root ZC sequences in SCE. However, the SCE performance
can be further enhanced since YL sequences are not all orthogonal.

Although efficient to implement, rectangularly-pulsed OFDM waveforms exhibit
large power spectral sidelobes due to discontinuity at pulse edges and thus
cause strong interference to adjacent channels \cite{CA1}, \cite{Adj channel
1}-\cite{Adj channel 3}. Specifically, rectangularly-pulsed OFDM waveforms
carrying ZC sequences have been shown to render widely spread waveform
spectrum with baseband spectral sidelobes decaying asymptotically as $f^{-2}$
\cite{preamble}-\cite{pilot2}. Although highly compact training waveform
spectrum can be composed by suppressing spectral sidelobes through delicate
signal processing techniques \cite{Adj channel 1}-\cite{Suppression 6}, the
feature of frequency-domain CA is altered in the transmitted waveform after
sidelobe suppression, thus compromising the performance characteristics of
initial synchronization, channel estimation, and SPI at the receiver. To
resolve the problem, several order-$I$ CA sequences have been recently
developed in \cite{preamble}-\cite{pilot2} to render extremely small
baseband power spectral sidelobes decaying asymptotically as $f^{-2I-2}$
with sidelobe-decaying (SD) order $I\geq 1$, and thus compose spectrally
compact training waveforms for robust fine initial synchronization \cite%
{preamble}-\cite{CA1} and accurate channel estimation \cite{CA1}-\cite%
{pilot2}. The larger the SD order $I$ is, the higher spectral compactness
the corresponding training waveform can achieve. Since frequency-domain CA
is sustained, order-$I$ CA sequences enable the same performance
characteristics as ZC sequences in initial synchronization and channel
estimation, while yielding much higher spectral compactness \cite{preamble}-%
\cite{pilot2}. In \cite{Wei pilot}-\cite{pilot2}, order-$I$ CA sequences $%
\mathcal{G}_{I}$ and $\mathcal{I}_{I}$ were first developed for a large
number of sequence lengths. For all composite and prime sequence lengths
larger than $11$, order-$I$ CA sequences $\widehat{\mathcal{G}}_{I}$ and $%
\widehat{\mathcal{I}}_{I}$ were further developed in \cite{CA1} and shown to
provide the SD order not smaller than order-$\widetilde{I}$ CA sequences $%
\mathcal{G}_{\widetilde{I}}$ and $\mathcal{I}_{\widetilde{I}}$. To meet the
needs of various SPI applications, four families containing mutually
orthogonal order-$I$ CA sequences were also developed in \cite{CA1} for
respective sequence types $\mathcal{G}_{I}$, $\mathcal{I}_{I}$, $\widehat{%
\mathcal{G}}_{I}$, and $\widehat{\mathcal{I}}_{I}$, based on the method of
phase model assigning (PMA), and denoted hereinafter by families $\mathcal{G}%
_{I}^{(\text{pma})}$, $\mathcal{I}_{I}^{(\text{pma})}$, $\widehat{\mathcal{G}%
}_{I}^{(\text{pma})}$, and $\widehat{\mathcal{I}}_{I}^{(\text{pma})}$ for
convenience. Nevertheless, the numbers of permissible orthogonal sequences
provided by these families are still insufficient for some SPI applications
requiring a large number of orthogonal CA sequences (like RA channel
identification and MIMO channel sounding) \cite{LTE}-\cite{5G}. This paper
is thus motivated to develop new families with an attempt to providing more
orthogonal order-$I$ CA sequences.

Based on the methods of degenerate PMA and augmented PMA, several modified
PMA sequence families are constructed herein to provide more orthogonal
order-$\widetilde{I}$ CA sequences ($\mathcal{G}_{\widetilde{I}}$, $\mathcal{%
I}_{\widetilde{I}}$, $\widehat{\mathcal{G}}_{\widetilde{I}}$, and $\widehat{%
\mathcal{I}}_{\widetilde{I}}$) than families $\mathcal{G}_{I}^{(\text{pma})}$%
, $\mathcal{I}_{I}^{(\text{pma})}$, $\widehat{\mathcal{G}}_{I}^{(\text{pma}%
)} $, and $\widehat{\mathcal{I}}_{I}^{(\text{pma})}$ by possibly trading off
the SD order $\widetilde{I}\leq I$. All developed order-$\widetilde{I}$
sequences can still provide much higher spectral compactness than ZC, YL,
and pseudorandom-noise (PN) sequences for the composed OFDM preamble/pilot
waveforms. Since sequences $\mathcal{I}_{\widetilde{I}}$ and $\widehat{%
\mathcal{I}}_{\widetilde{I}}$ can be similarly constructed like sequences $%
\mathcal{G}_{\widetilde{I}}$ and $\widehat{\mathcal{G}}_{\widetilde{I}}$,
only the new families composed of order-$\widetilde{I}$ CA sequences $%
\mathcal{G}_{\widetilde{I}}$ and $\widehat{\mathcal{G}}_{\widetilde{I}}$ are
elaborated in the following. The contribution of the paper is addressed as
follows.\footnote{\emph{Notations:} Boldface lower-case and upper-case
letters denote column vectors and matrices, respectively. Superscripts $t$, $%
\ast $, and $h$ denote transpose, complex conjugate, and conjugate
transpose, respectively. $\mathcal{Z}^{\ast }$, $\mathcal{Z}_{K}$ and $%
\mathcal{Z}_{K}^{+}$ are the set of nonnegative integers, $\{0,1,...,K-1\}$
and $\{1,2,...,K\}$, respectively. By default, $\mathcal{Z}_{0}^{+}$ is an
empty set. We also use $[x_{k};k\in \mathcal{Z}_{K}]$ to represent a $%
K\times 1$ vector with $x_{k}$ being the $k$-th entry, $\min \{x,y\}$ the
smaller between $x$\ and $y$, $((n))_{N}$ the modulo-$N$ value of $n$, $||%
\mathbf{x}||$ the Frobenius norm of vector $\mathbf{x}$, $\left\lceil
x\right\rceil $ the smallest integer that is not smaller than $x$, and $%
\left\lfloor x\right\rfloor $ the largest integer that is not larger than $x$%
. We let $\omega _{K}\triangleq \exp \{-j\frac{2\pi }{K}\}$ and denote $%
\mathbf{W}_{K}\triangleq \lbrack K^{-1/2}\omega _{K}^{mk};m\in \mathcal{Z}%
_{K},k\in \mathcal{Z}_{K}]$ as a $K\times K$ unitary DFT matrix with
normalized columns and rows. $\mathcal{E}\{\cdot \}$ denotes the expectation
operator. $j\triangleq \sqrt{-1}$ is the imaginary unit.}

\begin{itemize}
\item Degenerate PMA sequence families $\mathcal{G}_{\text{max,}\widetilde{I}%
}^{(\text{dpma,}\kappa )}$ and $\widetilde{\mathcal{G}}_{\text{max,}%
\widetilde{I}}^{(\text{dpma,}\kappa )}$ with sequence length $N$ are
constructed respectively under a proper level-$(\Omega (N)-\kappa )$
factorization of $N$ and under a near-proper level-$(\Omega (N)-\kappa )$
factorization of $N$ for $\kappa \in \mathcal{Z}_{\Omega (N)-1}^{+}$, where $%
\Omega (N)$ is the prime omega value of $N$ and denotes the multiplicity in
the prime factorization of $N$. Families $\mathcal{G}_{\text{max,}\widetilde{%
I}}^{(\text{dpma,}\kappa )}$ and $\widetilde{\mathcal{G}}_{\text{max,}%
\widetilde{I}}^{(\text{dpma,}\kappa )}$ can provide more orthogonal order-$%
\widetilde{I}$ CA sequences than the PMA sequence family $\mathcal{G}_{I}^{(%
\text{pma})}$, with or without trading off SD order $\widetilde{I}\leq I$.
When $\widetilde{\Omega }(N)>\Omega (N)$, degenerate PMA sequence families $%
\widehat{\mathcal{G}}_{\text{max,}\widetilde{I}}^{(\text{dpma,}\kappa )}$
are accordingly constructed under a combined proper level-$(\widetilde{%
\Omega }(N)-\kappa )$ factorization of $N$ for $\kappa \in \mathcal{Z}_{%
\widetilde{\Omega }(N)-1}^{+}$, where $\widetilde{\Omega }(N)$ is the
modified prime omega (MPO) value defined in \cite[eqs. 14-15]{CA1} and
denotes the increased multiplicity provided by all prime factorizations of
the properly decomposed values from $N$. Families $\widehat{\mathcal{G}}_{%
\text{max,}\widetilde{I}}^{(\text{dpma,}\kappa )}$ can provide more
orthogonal order-$\widetilde{I}$ CA sequences than the PMA sequence family $%
\widehat{\mathcal{G}}_{I}^{(\text{pma})}$, with or without trading off SD
order $\widetilde{I}\leq I$.

\item When $N$ meets $\widetilde{\Omega }(N)>\Omega (N)$, the augmented PMA
sequence family $\widehat{\mathcal{G}}_{I}^{(\text{apma})}$ is constructed
by virtue of phase-rotating every existing sequence in family $\widehat{%
\mathcal{G}}_{I}^{(\text{pma})}$ to generate more mutually orthogonal
sequence members, and thus provides double the number of orthogonal order-$I$
CA sequences in family $\widehat{\mathcal{G}}_{I}^{(\text{pma})}$ while
maintaining the same SD order. Based on the same phase-rotating method,
augmented degenerate PMA sequence family $\widehat{\mathcal{G}}_{\text{max,}%
\widetilde{I}}^{(\text{adpma,}\kappa )}$ is constructed from family $%
\widehat{\mathcal{G}}_{\text{max,}\widetilde{I}}^{(\text{dpma,}\kappa )}$
for a given $\kappa \in \mathcal{Z}_{\widetilde{\Omega }(N)-1}^{+}$ and
provides double the family size of $\widehat{\mathcal{G}}_{\text{max,}%
\widetilde{I}}^{(\text{dpma,}\kappa )}$ without trading off the SD order.

\item In comparison with ZC, YL, and PN sequence families, modified PMA
sequence families $\mathcal{G}_{\text{max,}\widetilde{I}}^{(\text{dpma,}%
\kappa )}$, $\widetilde{\mathcal{G}}_{\text{max,}\widetilde{I}}^{(\text{dpma,%
}\kappa )}$, $\widehat{\mathcal{G}}_{\text{max,}\widetilde{I}}^{(\text{dpma,}%
\kappa )}$, and $\widehat{\mathcal{G}}_{\text{max,}\widetilde{I}}^{(\text{%
adpma,}\kappa )}$ are demonstrated to enhance the performance
characteristics in uplink RA channel identification over indoor and urban
Rayleigh multipath environments exhibiting short-delay channel profiles,
thanks to the provision of more orthogonal CA sequences and thus the
mitigation of false identification. Meanwhile, the preamble waveforms
carrying order-$\widetilde{I}$ CA sequences from modified PMA sequence
families are attributed with much higher spectral compactness than those
carrying ZC, YL, and PN sequences.
\end{itemize}

The paper is organized as follows. Section II provides a review on order-$I$
CA sequences $\mathcal{G}_{I}$, $\widehat{\mathcal{G}}_{I}$, $\mathcal{I}%
_{I} $, and $\widehat{\mathcal{I}}_{I}$ \cite{CA1}-\cite{pilot2}. Section
III develops family $\mathcal{G}_{\text{max,}\widetilde{I}}^{(\text{dpma,}%
\kappa )}$ under a proper level-$(\Omega (N)-\kappa )$ factorization and
family $\widetilde{\mathcal{G}}_{\text{max,}\widetilde{I}}^{(\text{dpma,}%
\kappa )}$ under a near-proper level-$(\Omega (N)-\kappa )$ factorization,
both for $\kappa \in \mathcal{Z}_{\Omega (N)-1}^{+}$. When $\widetilde{%
\Omega }(N)>\Omega (N)$, family $\widehat{\mathcal{G}}_{I}^{(\text{apma})}$
is constructed in Section IV by the phase-rotating method. Families $%
\widehat{\mathcal{G}}_{\text{max,}\widetilde{I}}^{(\text{dpma,}\kappa )}$
and $\widehat{\mathcal{G}}_{\text{max,}\widetilde{I}}^{(\text{adpma,}\kappa
)}$ are also constructed under a combined proper level-$(\widetilde{\Omega }%
(N)-\kappa )$ factorization for $\kappa \in \mathcal{Z}_{\widetilde{\Omega }%
(N)-1}^{+}$. In Section V, the OFDM systems employing various CA sequence
families are compared for RA channel identification and spectral
compactness. Section VI concludes the paper.

\section{Order-$I$ Constant-Amplitude Sequences}

Consider the rectangularly-pulsed OFDM waveform carrying a sequence of $N$
complex symbols. In the nominal time interval of length $T$, these symbols
are modulated into $N$ uniformly-spaced subcarriers interleaved among $%
\gamma N$ subcarriers with a positive-integer-valued interleaving factor $%
\gamma $. The time interval is partitioned into a guard CP subinterval of
length $T_{\text{g}}$ followed by a useful signaling subinterval of length $%
T_{\text{d}}=T-T_{\text{g}}$, where $T_{\text{g}}=\alpha T_{\text{d}}$ and $%
\alpha $ is the guard ratio with $0<\alpha <1$. Denote $\mathbf{q}\triangleq %
\left[ q[n];n\in \mathcal{Z}_{N}\right] $ as the sequence in frequency
domain and $\widetilde{\mathbf{q}}\triangleq \left[ \tilde{q}[m];m\in 
\mathcal{Z}_{N}\right] =\mathbf{W}_{_{N}}^{h}\mathbf{q}$ as its inverse DFT
with $\left\Vert \mathbf{q}\right\Vert ^{2}=\left\Vert \widetilde{\mathbf{q}}%
\right\Vert ^{2}=1$. Throughout, $\mathbf{q}$ is restricted to have CA
symbols with $|q[n]|^{2}=1/N$, and thus its inverse DFT $\widetilde{\mathbf{q%
}}$ possesses the ZAC property, i.e., $\sum_{m\in \mathcal{Z}_{N}}%
\widetilde{q}[((m-n))_{N}](\widetilde{q}[m])^{\ast }=0$ for all integers $%
((n))_{N}\neq 0$ \cite{preamble}, \cite{CAZAC}.

Rectangularly-pulsed OFDM preamble/pilot waveforms are discontinuous if
identification symbols are not properly restricted and thus render large
baseband power spectral sidelobes decaying asymptotically as $f^{-2}$. In
practical OFDM systems, rectangularly pulsed preamble/pilot waveforms
carrying PN and ZC sequences render widely spread waveform spectrum with
baseband spectral sidelobes decaying asymptotically as $f^{-2}$ \cite%
{preamble}-\cite{pilot2}. By properly restricting identification symbols,
various order-$I$ CA sequences have been recently developed in \cite%
{preamble}-\cite{pilot2} to render extremely small baseband power spectral
sidelobes decaying asymptotically as $f^{-2I-2}$ and thus the corresponding
baseband power spectrum exhibits $I$-decaying sidelobes. Due to fast
sidelobe decaying, these order-$I$ CA sequences enhance the spectral
compactness of the corresponding OFDM preamble/pilot waveforms, while
achieving accurate channel estimation and robust fine initial time and
frequency synchronization owing to dual sequence properties of
frequency-domain CA and time-domain ZAC \cite{preamble}-\cite{CA1}.
Particularly in \cite{CA1}, four types of order-$I$ CA sequences $\mathcal{G}%
_{I}$, $\mathcal{I}_{I}$, $\widehat{\mathcal{G}}_{I}$, and $\widehat{%
\mathcal{I}}_{I}$ with sequence length $N$ have been developed in explicit
expressions for all composite sequence lengths and all prime sequence
lengths larger than $11$ under all parametric conditions on $\alpha \gamma $%
. In what follows, sequences $\mathcal{G}_{I}$, $\mathcal{I}_{I}$, $\widehat{%
\mathcal{G}}_{I}$, and $\widehat{\mathcal{I}}_{I}$ are briefly reviewed.

For convenience, an order-$I$ CA sequence $\mathbf{q}=[N^{\frac{-1}{2}%
}\left( -1\right) ^{n\gamma }\chi \left[ n\right] ;n\in \mathcal{Z}_{N}]$ is
described by a CA sequence $\bm{\chi }=[\chi \lbrack n];n\in \mathcal{Z}%
_{N}]$ with $\left\vert \chi \left[ n\right] \right\vert =1$ for all $n\in 
\mathcal{Z}_{N}$, and presented in two separate conditions, namely \emph{%
Condition A} that $\alpha \gamma $ is an integer and \emph{Condition B} that 
$\alpha \gamma $ is not an integer \cite{CA1}. Under \emph{Condition A}, if $%
\mathbf{q}$ satisfies%
\begin{eqnarray*}
\text{\emph{Constraint A:}}\mathit{\ }\bm{\mu }_{\beta }^{t}\bm{\chi 
}=0\text{ for all }\beta \in \mathcal{Z}_{I}\text{ but }\bm{\mu }_{I}^{t}%
\bm{\chi }\neq 0
\end{eqnarray*}%
for a positive integer $I\in \mathcal{Z}_{N-1}^{+}$ where $\bm{\mu }%
_{\beta }\triangleq \lbrack n^{\beta };n\in \mathcal{Z}_{N}]$, the
corresponding baseband power spectrum exhibits $I$-decaying sidelobes. Under 
\emph{Condition B}, if $\mathbf{q}$ satisfies%
\begin{eqnarray*}
&&\text{\emph{Constraint B:}\textit{\ }}\bm{\mu }_{\beta }^{t}\bm{%
\chi }=0\text{ and }\widetilde{\bm{\mu }}_{\beta }^{t}\bm{\chi }=0%
\text{ for all }\beta \in \mathcal{Z}_{I} \\
&&\qquad \qquad \qquad \text{but }\bm{\mu }_{I}^{t}\bm{\chi }\neq 0%
\text{ or }\widetilde{\bm{\mu }}_{I}^{t}\bm{\chi }\neq 0
\end{eqnarray*}%
for a positive integer $I\in \mathcal{Z}_{\lfloor (N-1)/2\rfloor }^{+}$
where $\widetilde{\bm{\mu }}_{\beta }\triangleq \lbrack e^{-j2\pi
n\alpha \gamma }n^{\beta };n\in \mathcal{Z}_{N}]$, the corresponding
baseband power spectrum exhibits $I$-decaying sidelobes. Throughout, we
consider the prime factorization $N=\prod\nolimits_{m=0}^{\Omega
(N)-1}P_{m} $ where prime integers $P_{m}$ may not be all distinct. Due to
the constraints, the largest possible family size $\Psi _{\text{max}}(N)$ is
limited by $\Psi _{\text{max}}(N)=N-I$ under \emph{Condition A} and $\Psi _{%
\text{max}}(N)=N-2I$ under \emph{Condition B}, for any sequence family
containing mutually orthogonal sequences of length $N$.

\emph{A) Sequence }$\mathcal{G}_{I}$\emph{: }Arrange prime factors $P_{m}$
in descending order $P_{0}\geq P_{1}\geq ...\geq P_{\Omega (N)-1}$. Define $%
\phi _{m}\triangleq \prod\nolimits_{k=0}^{m-1}P_{k}$ for $m\in \mathcal{Z}%
_{\Omega (N)-1}^{+}$\ and $\phi _{0}=1$. An order-$I$ CA sequence $\mathcal{G%
}_{I}$ is described as%
\begin{equation}
\chi \lbrack \sum\nolimits_{m\in \mathcal{Z}_{\Omega (N)}}l_{m}\phi
_{m}]=\exp \{j\sum\nolimits_{m\in \mathcal{Z}_{\Omega (N)}}\theta
_{m}[l_{m}]\}  \label{2}
\end{equation}%
for all $l_{0}\in \mathcal{Z}_{P_{0}}$, $l_{1}\in \mathcal{Z}_{P_{1}}$,..., $%
l_{\Omega (N)-1}\in \mathcal{Z}_{P_{\Omega (N)-1}}$ under \emph{Condition A}%
, and%
\begin{eqnarray}
&&\chi \lbrack \sum\nolimits_{m\in \mathcal{Z}_{\Omega (N)}}l_{m}\phi
_{m}]=\exp \{j\sum\nolimits_{m\in \mathcal{Z}_{\Omega (N)}}\theta
_{m}[l_{m}]  \label{3} \\
&&\qquad \qquad \qquad +j2\pi \alpha \gamma \sum\nolimits_{n\in \mathcal{Z}%
_{\left\lfloor \Omega (N)/2\right\rfloor }}l_{2n+1}\phi _{2n+1}\}  \notag
\end{eqnarray}%
for all $l_{0}\in \mathcal{Z}_{P_{0}}$, $l_{1}\in \mathcal{Z}_{P_{1}}$,..., $%
l_{\Omega (N)-1}\in \mathcal{Z}_{P_{\Omega (N)-1}}$ under \emph{Condition B}%
. Here, the phases $\theta _{m}[l_{m}]$ are restricted by%
\begin{equation}
\sum\nolimits_{l_{m}\in \mathcal{Z}_{P_{m}}}\exp \{j\theta _{m}[l_{m}]\}=0%
\text{ for all }m\in \mathcal{Z}_{\Omega (N)}.  \label{restriction 1}
\end{equation}%
For a given $N$, sequence $\mathcal{G}_{I}$ yields the SD order $I\geq
\Omega (N)$ under \emph{Condition A }and $I\geq \left\lfloor \Omega
(N)/2\right\rfloor $ under \emph{Condition B}.

Orthogonal sequence family $\mathcal{G}_{I}^{(\text{pma})}$ have been
obtained from the PMA method in \cite[Subsection III\emph{.B}]{CA1}. For a
given index vector $\bm{\nu }=[v_{m};m\in \mathcal{Z}_{\Omega (N)}]$
with $v_{m}\in \mathcal{Z}_{P_{m}-1}^{+}$ for all $m\in \mathcal{Z}_{\Omega
(N)}$, a sequence in family $\mathcal{G}_{I}^{(\text{pma})}$ can be uniquely
specified by $\bm{\nu }$ and formed by assigning%
\begin{equation}
\theta _{m}\left[ l_{m}\right] =\frac{2\pi \nu _{m}l_{m}}{P_{m}}\text{ for
all }l_{m}\in \mathcal{Z}_{P_{m}}\text{ and }m\in \mathcal{Z}_{\Omega (N)}
\label{4}
\end{equation}%
under either \emph{Condition A }or \emph{Condition B}. By varying $\bm{%
\nu }$ exclusively, family $\mathcal{G}_{I}^{(\text{pma})}$ can be
constructed accordingly and it contains $\Psi (N)\triangleq
\prod\nolimits_{m=0}^{\Omega (N)-1}(P_{m}-1)$ orthogonal order-$I$ CA
sequences. Apparently, all order-$I$ sequences in $\mathcal{G}_{I}^{(\text{%
pma})}$ are mutually orthogonal, i.e., $\mathbf{q}_{l}^{h}\mathbf{q}_{k}=%
\widetilde{\mathbf{q}}_{l}^{h}\widetilde{\mathbf{q}}_{k}=0$ for any two
different sequences $\mathbf{q}_{l}$ and $\mathbf{q}_{k}$ in $\mathcal{G}%
_{I}^{(\text{pma})}$.

Consider a leader sequence $\mathbf{q}_{\text{lead}}$ in $\mathcal{G}_{I}^{(%
\text{pma})}$, specified by $\bm{\nu }=[v_{0},v_{1},...,v_{\Omega
(N)-1}]^{t}$. Denote $\widetilde{\mathbf{q}}_{\text{lead}}^{(k)}=\left[ 
\widetilde{q}_{\text{lead}}\left[ ((i+k))_{N}\right] ;i\in \mathcal{Z}_{N}%
\right] $ as the $k$-cyclically-shifted version of $\widetilde{\mathbf{q}}_{%
\text{lead}}$ (i.e., the inverse DFT of $\mathbf{q}_{\text{lead}}$)\ and $%
\mathbf{q}_{\text{lead}}^{(k)}=[q_{\text{lead}}[n]\exp \{j2\pi nk/N\};n\in 
\mathcal{Z}_{N}]$ as its DFT. According to \cite{CA1}, the set of admissible
cyclic shifts for which $\mathbf{q}_{\text{lead}}^{(k)}$ is still an order-$%
I $ CA sequence in $\mathcal{G}_{I}^{(\text{pma})}$ is specified by $%
\mathcal{U}(\nu _{0})\triangleq \{lN/P_{\max }|l\in \mathcal{Z}_{P_{\max }}$
but $l\neq P_{\max }-v_{0}\}$ where $P_{\max }\triangleq \max_{m\in \mathcal{%
Z}_{\Omega (N)}}P_{m}$ is the largest prime factor and $v_{0}$ is the
leading entry in $\bm{\nu }$. From $\mathbf{q}_{\text{lead}}$, we can
thus specify the cyclic-shift (CS) CA sequence subfamily $\mathcal{G}_{I}^{(%
\text{cs})}(\mathbf{q}_{\text{lead}})$ which contains all
cyclically-shiftable order-$I$ CA sequences obtained by cyclically shifting $%
\widetilde{\mathbf{q}}_{\text{lead}}$ with shifts in $\mathcal{U}(\nu _{0})$%
, as%
\begin{equation}
\mathcal{G}_{I}^{(\text{cs})}(\mathbf{q}_{\text{lead}})\triangleq \{\mathbf{q%
}_{\text{lead}}^{(k)}|k\in \mathcal{U}(\nu _{0})\}\text{ if }\mathbf{q}_{%
\text{lead}}\bm{\in }\mathcal{G}_{I}^{(\text{pma})}  \label{5}
\end{equation}%
under either \emph{Condition A }or \emph{Condition B}. The factor $N/P_{\max
}$ defining $\mathcal{U}(\nu _{0})$ is the family CSD for generating
cyclically-shiftable CA sequence subfamily $\mathcal{G}_{I}^{(\text{cs})}(%
\mathbf{q}_{\text{lead}})$. Notably, $\mathcal{G}_{I}^{(\text{cs})}(\mathbf{q%
}_{\text{lead}})$ contains $P_{\max }-1$ different sequences in $\mathcal{G}%
_{I}^{(\text{pma})}$, which are specified by identical indices $%
v_{1},v_{2},...,v_{\Omega (N)-1}$. Therefore, by varying $%
v_{1},v_{2},...,v_{\Omega (N)-1}$, we can obtain $\Psi (N)/(P_{\max }-1)$
mutually exclusive subfamilies $\mathcal{G}_{I}^{(\text{cs})}(\mathbf{q}_{%
\text{lead}})$ constructed from all permissible subfamily leaders $\mathbf{q}%
_{\text{lead}}$ specified by different index subvectors $%
[v_{1},v_{2},...,v_{\Omega (N)-1}]^{t}$. In $\mathcal{G}_{I}^{(\text{cs})}(%
\mathbf{q}_{\text{lead}})$, all orthogonal order-$I$ CA sequences can be
easily obtained by cyclically shifting the inverse DFT of a subfamily leader 
$\mathbf{q}_{\text{lead}}$.

\emph{B) Sequence }$\mathcal{I}_{I}$\emph{: }Arrange prime factors $P_{m}$
in ascending order $P_{0}\leq P_{1}\leq ...\leq P_{\Omega (N)-1}$. Define $%
\psi _{m}=N/\phi _{m+1}$\ for $m\in \mathcal{Z}_{\Omega (N)-1}$ and $\psi
_{\Omega (N)-1}=1$. An order-$I$ CA sequence $\mathcal{I}_{I}$ is defined
similarly to sequence $\mathcal{G}_{I}$ as in (\ref{2})-(\ref{4}) with $\phi
_{m}\rightarrow \psi _{m}$ for $m\in \mathcal{Z}_{\Omega (N)}$. With
sequence length $N$, sequence $\mathcal{I}_{I}$ yields the SD order $I\geq
\Omega (N)$ under \emph{Condition A }and $I\geq \left\lfloor \Omega
(N)/2\right\rfloor $ under \emph{Condition B}. By varying the index vector $%
\bm{\nu }$ exclusively, the orthogonal sequence family $\mathcal{I}%
_{I}^{(\text{pma})}$ can be likewise constructed and it contains $\Psi (N)$
mutually orthogonal order-$I$ CA sequences. For a given $\mathbf{q}_{\text{%
lead}}$ in $\mathcal{I}_{I}^{(\text{pma})}$ specified by $\bm{\nu }%
=[v_{0},v_{1},...,v_{\Omega (N)-1}]$, the cyclic-shift sequence subfamily $%
\mathcal{I}_{I}^{(\text{cs})}(\mathbf{q}_{\text{lead}})$ can be obtained as%
\begin{equation}
\mathcal{I}_{I}^{(\text{cs})}(\mathbf{q}_{\text{lead}})\triangleq \{\mathbf{q%
}_{\text{lead}}^{(k)}|k\in \mathcal{U}(\nu _{\Omega (N)-1})\}\text{ if }%
\mathbf{q}_{\text{lead}}\bm{\in }\mathcal{I}_{I}^{(\text{pma})}
\end{equation}%
under either \emph{Condition A }or \emph{Condition B}. Thus, $\mathcal{I}%
_{I}^{(\text{cs})}(\mathbf{q}_{\text{lead}})$ contains $P_{\max }-1$
different sequences in $\mathcal{I}_{I}^{(\text{pma})}$, which are specified
by identical indices $v_{0},v_{1},...,v_{\Omega (N)-2}$. We can obtain $\Psi
(N)/(P_{\max }-1)$ mutually exclusive cyclically-shiftable CA sequence
subfamilies $\mathcal{I}_{I}^{(\text{cs})}(\mathbf{q}_{\text{lead}})$
constructed from all permissible subfamily leaders $\mathbf{q}_{\text{lead}}$
specified by different index subvectors $[v_{0},v_{1},...,v_{\Omega
(N)-2}]^{t}$.

Due to the similarity between $\mathcal{G}_{I}^{(\text{pma})}$ and $\mathcal{%
I}_{I}^{(\text{pma})}$, only modified PMA families from $\mathcal{G}_{I}^{(%
\text{pma})}$ are elaborated below.

\emph{C) Sequences }$\widehat{\mathcal{G}}_{I}$\emph{\ and }$\widehat{%
\mathcal{I}}_{I}$\emph{:} For a given sequence length $N$, order-$I$ CA
sequences $\widehat{\mathcal{G}}_{I}$ and $\widehat{\mathcal{I}}_{I}$ are
constructed from concatenating component CA subsequences with shorter
lengths as follows. First, $N$ is properly decomposed into $N=\sum_{\rho \in 
\mathcal{Z}_{L}}\widetilde{N}^{(\rho )}$ where $\widetilde{N}^{(\rho
)}=\prod\nolimits_{m=0}^{\Omega (\widetilde{N}^{(\rho )})-1}P_{m}^{(\rho )}$
with prime factors $P_{0}^{(\rho )}\geq P_{1}^{(\rho )}\geq ...\geq
P_{\Omega (\widetilde{N}^{(\rho )})-1}^{(\rho )}$ arranged for all $\rho \in 
\mathcal{Z}_{L}$, and $\bm{\chi }$ is accordingly partitioned into $L$
subsequences $\bm{\chi }_{0},\bm{\chi }_{1},...,\bm{\chi }_{L-1}$
of lengths $\widetilde{N}^{(0)}$, $\widetilde{N}^{(1)}$,..., $\widetilde{N}%
^{(L-1)}$, respectively, i.e., $\bm{\chi }=\left[ \bm{\chi }_{0}^{t},%
\bm{\chi }_{1}^{t},...,\bm{\chi }_{L-1}^{t}\right] ^{t}$. Second,
subsequences $\bm{\chi }_{0},\bm{\chi }_{1},...,\bm{\chi }_{L-1}$
are constructed in the forms (\ref{2})-(\ref{4}) with $\phi _{m}\rightarrow
\phi _{m}^{(\rho )}$ for $m\in \mathcal{Z}_{\Omega (\widetilde{N}^{(\rho
)})} $ and $\rho \in \mathcal{Z}_{L}$, and then concatenated to form
sequence $\widehat{\mathcal{G}}_{I}$. Sequence $\widehat{\mathcal{I}}_{I}$
is formed similarly with subsequences constructed in the forms (\ref{2})-(%
\ref{4}) for all $\rho \in \mathcal{Z}_{L}$ with $\phi _{m}\rightarrow \psi
_{m}^{(\rho )} $ for $m\in \mathcal{Z}_{\Omega (\widetilde{N}^{(\rho )})}$
and $P_{0}^{(\rho )}\leq P_{1}^{(\rho )}\leq ...\leq P_{\Omega (\widetilde{N}%
^{(\rho )})-1}^{(\rho )}$ rearranged.

With a proper decomposition $N=\sum_{\rho \in \mathcal{Z}_{L}}\widetilde{N}%
^{(\rho )}$, order-$I$ CA sequences $\widehat{\mathcal{G}}_{I}$ and $%
\widehat{\mathcal{I}}_{I}$ yield the SD order $I\geq \widetilde{\Omega }(N)$
under \emph{Condition A }and $I\geq \left\lfloor \widetilde{\Omega }%
(N)/2\right\rfloor $ under \emph{Condition B}, where the \emph{proper}
decomposition $\{\widetilde{N}^{(\rho )};\rho \in \mathcal{Z}_{L}\}$ can
achieve the MPO value $\widetilde{\Omega }(N)$, defined in \cite[eqs. 14-15]%
{CA1} as%
\begin{equation}
\widetilde{\Omega }(N)=\max\limits_{L\in \mathcal{Z}_{\left\lfloor
N/2\right\rfloor }^{+}}\max\limits_{\substack{ \widetilde{N}^{(0)}\geq 
\widetilde{N}^{(1)}\geq ...\geq \widetilde{N}^{(L-1)}\geq 2  \\ \widetilde{N}%
^{(0)}+\widetilde{N}^{(1)}+...+\widetilde{N}^{(L-1)}=N}}\min\limits_{\rho
\in \mathcal{Z}_{L}}\Omega (\widetilde{N}^{(\rho )}).  \label{MPO}
\end{equation}%
Notably, the proper decomposition is not necessarily unique for arbitrary
lengths $N$ and can assure $\widetilde{\Omega }(N)\geq \Omega (N)$.
Particularly, $\widetilde{\Omega }(N)>\Omega (N)$ is guaranteed if and only
if (iff) $N$ is not any one of the following forms 
\begin{eqnarray}
N &=&2^{a}\times 3^{b}\times 5^{c}  \label{8} \\
N &=&2^{a}\times 3^{b}\times 7^{d}  \label{9} \\
N &=&2^{a}\times 3^{b}\times 11^{e}  \label{10}
\end{eqnarray}%
where the nature numbers $a$, $b$, $c$, $d$ and $e$ are restricted to $a$, $%
b\in \mathcal{Z}^{\ast }$, $c\in \mathcal{Z}_{4}$, $d\in \mathcal{Z}_{3}$,
and $e\in \mathcal{Z}_{2}$ \cite[\emph{Property 5}]{CA1}. In the case, order-%
$I$ CA sequences $\widehat{\mathcal{G}}_{I}$ and $\widehat{\mathcal{I}}_{I}$
can yield higher SD order than order-$\widetilde{I}$ CA sequences $\mathcal{G%
}_{\widetilde{I}}$ and $\mathcal{I}_{\widetilde{I}}$. Conversely, when $N$
is one of the above three forms, sequences $\mathcal{G}_{\widetilde{I}}$ and 
$\mathcal{I}_{\widetilde{I}}$ can provide comparable SD order to sequences $%
\widehat{\mathcal{G}}_{I}$ and $\widehat{\mathcal{I}}_{I}$ due to $%
\widetilde{\Omega }(N)=\Omega (N)$. For a given $N$, a proper decomposition
and the associated $\widetilde{\Omega }(N)$ can be efficiently sought from 
\emph{Procedure 1} and \emph{Property 4} in \cite{CA1}, where some examples
for medium and large $N$ values are also listed in \cite[Tables I and II]%
{CA1}.

Orthogonal sequence family $\widehat{\mathcal{G}}_{I}^{(\text{pma})}$ has
also been obtained from the PMA method \cite{CA1}. Consider the prime
factorizations $\widetilde{N}^{(\rho )}=\prod\nolimits_{m=0}^{\Omega (%
\widetilde{N}^{(\rho )})-1}P_{m}^{(\rho )}$ for all $\rho \in \mathcal{Z}%
_{L} $. For the given index vectors $\bm{\nu }^{(\rho )}=[v_{m}^{(\rho
)};m\in \mathcal{Z}_{\Omega (\widetilde{N}^{(\rho )})}]$ with $v_{m}^{(\rho
)}\in \mathcal{Z}_{P_{m}^{(\rho )}-1}^{+}$ for all $\rho \in \mathcal{Z}_{L}$%
, all subsequences of the corresponding sequence $\widehat{\mathcal{G}}_{I}$
in family $\widehat{\mathcal{G}}_{I}^{(\text{pma})}$ can be thus specified
by these $\bm{\nu }^{(\rho )}$ and formed from the phase assignment in (%
\ref{4}). By varying $\bm{\nu }^{(\rho )}$ exclusively and concurrently
for all $\rho \in \mathcal{Z}_{L}$, family $\widehat{\mathcal{G}}_{I}^{(%
\text{pma})}$ is constructed accordingly and it contains $\widehat{\Psi }%
(N)\triangleq \min_{\rho \in \mathcal{Z}_{L}}\Psi (\widetilde{N}^{(\rho )})$
orthogonal order-$I$ CA sequences, where $\Psi (\widetilde{N}^{(\rho
)})=\prod\nolimits_{m=0}^{\Omega (\widetilde{N}^{(\rho )})-1}(P_{m}^{(\rho
)}-1)$ for all $\rho \in \mathcal{Z}_{L}$. Notably, the orthogonal sequence
family $\widehat{\mathcal{G}}_{I}^{(\text{pma})}$ is not necessarily unique
for a given length $N$ since the proper decomposition for $N$ is not
necessarily unique. When $N$ is not any one of the forms in (\ref{8})-(\ref%
{10}), any orthogonal sequence in family $\widehat{\mathcal{G}}_{I}^{(\text{%
pma})}$ can not be obtained from cyclically shifting the inverse DFT of
another sequence in the family, due to the proper decomposition of $N$ \cite%
{CA1}. Without limitation by minimum CSD, the latter feature permits the use
of all orthogonal sequences in such family $\widehat{\mathcal{G}}_{I}^{(%
\text{pma})}$ and its modified families (to be developed below) for SPI
applications in the uplink cellular environment.

The orthogonal sequence family $\widehat{\mathcal{I}}_{I}^{(\text{pma})}$ is
likewise constructed. Due to the similarity between $\widehat{\mathcal{G}}%
_{I}^{(\text{pma})}$ and $\widehat{\mathcal{I}}_{I}^{(\text{pma})}$, only
modified PMA families from $\widehat{\mathcal{G}}_{I}^{(\text{pma})}$ are
elaborated herein.

When the sequence length $N$ is not any one of the forms in (\ref{8})-(\ref%
{10}) and has $\widetilde{\Omega }(N)>\Omega (N)$, we can construct family $%
\widehat{\mathcal{G}}_{I}^{(\text{pma})}$ with larger SD order than family $%
\mathcal{G}_{\widetilde{I}}^{(\text{pma})}$. Different from family $\mathcal{%
G}_{\widetilde{I}}^{(\text{pma})}$, all sequences in family $\widehat{%
\mathcal{G}}_{I}^{(\text{pma})}$ can not be obtained through cyclically
shifting the inverse DFT of another sequence. Conversely, when $N$ follows
any one of the forms in (\ref{8})-(\ref{10}) and exhibits $\widetilde{\Omega 
}(N)=\Omega (N)$, family $\mathcal{G}_{\widetilde{I}}^{(\text{pma})}$ can
yield comparable SD order to family $\widehat{\mathcal{G}}_{I}^{(\text{pma}%
)} $ and is composed of $\Psi (N)/(P_{\max }-1)$ mutually exclusive
subfamilies $\mathcal{G}_{\widetilde{I}}^{(\text{cs})}(\mathbf{q}_{\text{lead%
}})$ for all permissible subfamily leaders $\mathbf{q}_{\text{lead}}$, as
shown in Subsection II.\emph{A}. Each subfamily $\mathcal{G}_{\widetilde{I}%
}^{(\text{cs})}(\mathbf{q}_{\text{lead}})$ contains $P_{\max }-1$ sequences
generated by cyclically shifting the inverse DFT of a subfamily leader $%
\mathbf{q}_{\text{lead}}$ in family $\mathcal{G}_{\widetilde{I}}^{(\text{pma}%
)}$\ with family CSD $N/P_{\max }$.

The following sections are devoted to the development of two types of new
orthogonal sequence families, namely degenerate PMA sequence families $%
\mathcal{G}_{\widetilde{I}}^{(\text{dpma,}\kappa )}$, $\widehat{\mathcal{G}}%
_{\widetilde{I}}^{(\text{dpma,}\kappa )}$ and augmented PMA sequence
families $\widehat{\mathcal{G}}_{I}^{(\text{apma})}$, $\widehat{\mathcal{G}}%
_{\widetilde{I}}^{(\text{adpma,}\kappa )}$. For a composite length $N$,
family $\mathcal{G}_{\widetilde{I}}^{(\text{dpma,}\kappa )}$ contains
orthogonal order-$\widetilde{I}$ CA sequences $\mathcal{G}_{\widetilde{I}}$
with the larger family size than family $\mathcal{G}_{I}^{(\text{pma})}$ by
sacrificing the SD order in some cases. When $N$ meets $\widetilde{\Omega }%
(N)>\Omega (N)$, family $\widehat{\mathcal{G}}_{I}^{(\text{apma})}$ exhibits
double the family size as family $\widehat{\mathcal{G}}_{I}^{(\text{pma})}$
while maintaining the same SD order. Moreover, degenerate PMA sequence
family $\widehat{\mathcal{G}}_{\widetilde{I}}^{(\text{dpma,}\kappa )}$ and
augmented degenerate PMA sequence family $\widehat{\mathcal{G}}_{\widetilde{I%
}}^{(\text{adpma,}\kappa )}$ are also developed from degenerating families $%
\widehat{\mathcal{G}}_{I}^{(\text{pma})}$ and $\widehat{\mathcal{G}}_{I}^{(%
\text{apma})}$, respectively, by trading off the SD order.

\section{Families $\mathcal{G}_{\rm{max,}\protect\widetilde{\emph{I}}}^{(\rm{%
dpma,}\protect\kappa )}$ and $\protect\widetilde{\mathcal{G}}_{\rm{max,}%
\protect\widetilde{\emph{I}}}^{(\rm{dpma,}\protect\kappa )}$}

Consider a composite length $N$ with the prime factorization $%
N=\prod\nolimits_{m=0}^{\Omega (N)-1}P_{m}$ and $\Omega (N)>2$. With a
given $\kappa \in \mathcal{Z}_{\Omega (N)-1}^{+}$, many families $\mathcal{G}%
_{\widetilde{I}}^{(\text{dpma,}\kappa )}$ can be degenerated from family $%
\mathcal{G}_{I}^{(\text{pma})}$ with identical or less SD order. Based on a
particular level-$(\Omega (N)-\kappa )$ factorization $N=\prod%
\nolimits_{m=0}^{\Omega (N)-\kappa -1}A_{m}$ where factors $A_{m}$ may not
be all primes and are arranged in descending order, a family $\mathcal{G}_{%
\widetilde{I}}^{(\text{dpma,}\kappa )}$ can be constructed by the same PMA
method constructing family $\mathcal{G}_{I}^{(\text{pma})}$. Specifically,
such family $\mathcal{G}_{\widetilde{I}}^{(\text{dpma,}\kappa )}$ contains $%
\prod\nolimits_{m=0}^{\Omega (N)-\kappa -1}(A_{m}-1)$ orthogonal order-$%
\widetilde{I}$ CA sequences $\mathcal{G}_{\widetilde{I}}$ by varying the
index vector $\bm{\nu }=[v_{m};m\in \mathcal{Z}_{\Omega (N)-\kappa }]$
exclusively with $v_{m}\in \mathcal{Z}_{A_{m}-1}^{+}$ for all $m\in \mathcal{%
Z}_{\Omega (N)-\kappa }$, and exhibits the SD order $\widetilde{I}\geq
\Omega (N)-\kappa $ under \emph{Condition A} and $\widetilde{I}\geq
\left\lfloor (\Omega (N)-\kappa )/2\right\rfloor $ under \emph{Condition B}.
As $\Omega (N)$ is odd, any family $\mathcal{G}_{\widetilde{I}}^{(\text{dpma,%
}1)}$ based on any level-$(\Omega (N)-1)$ factorization yields the SD order $%
\widetilde{I}\geq $ $\left\lfloor \Omega (N)/2\right\rfloor $ under \emph{%
Condition B}, which may exhibit the same SD as family $\mathcal{G}_{I}^{(%
\text{pma})}$ based on the prime factorization.

For a fixed $\kappa \in \mathcal{Z}_{\Omega (N)-2}^{+}$, the family sizes
for different families $\mathcal{G}_{\widetilde{I}}^{(\text{dpma,}\kappa )}$
are not necessarily identical, depending on corresponding level-$(\Omega
(N)-\kappa )$ factorizations. Besides, more orthogonal PMA sequences can be
obtained by using a larger $\kappa $ in $\mathcal{Z}_{\Omega (N)-1}^{+}$ in
that $A_{m}A_{n}-1$ is strictly larger than $(A_{m}-1)(A_{n}-1)$ for any two
integer factors with $A_{m},A_{n}>1$. The largest family size $\Psi _{\text{%
max}}(N)=N-1$ under \emph{Condition A} is exactly achieved by the only
family $\mathcal{G}_{\widetilde{I}}^{(\text{dpma,}\Omega (N)-1)}$ based on
the level-$1$ factorization $N=A_{0}$. For a fixed $\kappa \in \mathcal{Z}%
_{\Omega (N)-2}^{+}$, a particular level-$(\Omega (N)-\kappa )$
factorization $N=\prod\nolimits_{m=0}^{\Omega (N)-\kappa -1}A_{m}^{(\kappa
)}$ is said to be a \emph{proper} level-$(\Omega (N)-\kappa )$ factorization
if $\Psi ^{(\text{dpma,}\kappa )}(N)\triangleq \prod\nolimits_{m=0}^{\Omega
(N)-\kappa -1}(A_{m}^{(\kappa )}-1)$ is the achievable largest family size
among all possible families $\mathcal{G}_{\widetilde{I}}^{(\text{dpma,}%
\kappa )}$. Such proper level-$(\Omega (N)-\kappa )$ factorization may not
be unique. Under a proper factorization, the corresponding family $\mathcal{G%
}_{\widetilde{I}}^{(\text{dpma,}\kappa )}$ is dubbed $\mathcal{G}_{\text{max,%
}\widetilde{I}}^{(\text{dpma,}\kappa )}$ for notational convenience. Family $%
\mathcal{G}_{\text{max,}\widetilde{I}}^{(\text{dpma,}\kappa )}$ consists of $%
\Psi ^{(\text{dpma,}\kappa )}(N)/(A_{\max }^{(\kappa )}-1)$ mutually
exclusive CS sequence subfamilies $\mathcal{G}_{\text{max,}\widetilde{I}}^{(%
\text{cs,}\kappa )}(\mathbf{q}_{\text{lead}})$ for all permissible subfamily
leaders $\mathbf{q}_{\text{lead}}$, where $A_{\max }^{(\kappa )}\triangleq
\max_{m\in \mathcal{Z}_{\Omega (N)-\kappa }}A_{m}^{(\kappa )}$. Each
subfamily $\mathcal{G}_{\text{max,}\widetilde{I}}^{(\text{cs,}\kappa )}(%
\mathbf{q}_{\text{lead}})$ contains $A_{\max }^{(\kappa )}-1$ sequences
generated by cyclically shifting the inverse DFT of a subfamily leader $%
\mathbf{q}_{\text{lead}}$ in family $\mathcal{G}_{\text{max,}\widetilde{I}%
}^{(\text{dpma,}\kappa )}$ with family CSD $\varpi _{\mathcal{G}}^{(\kappa
)}\triangleq N/A_{\max }^{(\kappa )}$, Below, proper level-$(\Omega
(N)-\kappa )$ factorizations with $\kappa =1$ and $\kappa =2$ are first
developed in closed-form expressions. An exclusive search procedure is then
proposed to find proper level-$(\Omega (N)-\kappa )$ factorizations with all 
$\kappa \in \mathcal{Z}_{\Omega (N)-2}^{+}$. Last, for sequence lengths $N$
with\ $\Omega (N)>4$, near-proper level-$(\Omega (N)-\kappa )$
factorizations $N=\prod\nolimits_{m=0}^{\Omega (N)-\kappa -1}\widetilde{A}%
_{m}^{(\kappa )}$ for all $\kappa \in \mathcal{Z}_{\Omega (N)-2}^{+}-%
\mathcal{Z}_{2}^{+}$ are presented in closed-form expressions to construct
another degenerate PMA sequence family $\widetilde{\mathcal{G}}_{\text{max,}%
\widetilde{I}}^{(\text{dpma,}\kappa )}$, which also gives a larger family
size than $\mathcal{G}_{I}^{(\text{pma})}$.

For presentation convenience, prime factors in $N=\prod\nolimits_{m=0}^{%
\Omega (N)-1}P_{m}$ are arranged below in ascending order $P_{0}\leq
P_{1}\leq ...\leq P_{\Omega (N)-1}$ for the development of proper and
near-proper level-$(\Omega (N)-\kappa )$ factorizations and the developed
factors in $N=\prod\nolimits_{m=0}^{\Omega (N)-\kappa -1}A_{m}^{(\kappa )}$
and $N=\prod\nolimits_{m=0}^{\Omega (N)-\kappa -1}\widetilde{A}%
_{m}^{(\kappa )}$ are not arranged in any order. Notably, to construct order-%
$\widetilde{I}$ CA sequences in families $\mathcal{G}_{\text{max,}\widetilde{%
I}}^{(\text{dpma,}\kappa )}$ and $\widetilde{\mathcal{G}}_{\text{max,}%
\widetilde{I}}^{(\text{dpma,}\kappa )}$, the developed factors have to be
rearranged beforehand in descending order, i.e., $A_{0}^{(\kappa )}\geq
A_{1}^{(\kappa )}\geq ...\geq A_{\Omega (N)-\kappa -1}^{(\kappa )}$ and $%
\widetilde{A}_{0}^{(\kappa )}\geq \widetilde{A}_{1}^{(\kappa )}\geq ...\geq 
\widetilde{A}_{\Omega (N)-\kappa -1}^{(\kappa )}$.

\emph{A) Proper Level-}$(\Omega (N)-1)$\emph{\ Factorization for }$\Omega
(N)>2$\emph{: }With $N=\prod\nolimits_{m=0}^{\Omega (N)-1}P_{m}$ and $%
\Omega (N)>2$, $N$ can be factorized into $\Omega (N)-1$ factors only when
two specific prime factors $P_{i}$ and $P_{n}$ are chosen from $\{P_{m};m\in 
\mathcal{Z}_{\Omega (N)}\}$ and merged into one composite factor $P_{i}P_{n}$%
. Under such factorization, one family $\mathcal{G}_{\widetilde{I}}^{(\text{%
dpma,}1)}$ can be formed with the family size $\Psi (N)\times
f([P_{i},P_{n}])$, where the function $f(\mathbf{a}^{t})$ is defined by%
\begin{equation}
f(\mathbf{a}^{t})=\frac{\prod\nolimits_{m\in \mathcal{Z}_{M}}a_{m}-1}{%
\prod\nolimits_{m\in \mathcal{Z}_{M}}(a_{m}-1)}  \label{11}
\end{equation}%
with $\mathbf{a}=[a_{m};m\in \mathcal{Z}_{M}]$ being an $M$-tuple argument
with all integer-valued entries $a_{m}>1$. This family size can be maximized
by choosing $P_{i}$ and $P_{n}$ properly based on \emph{Lemma 1}, which is
proven in \emph{Appendix A}.

\emph{Lemma 1:} Consider two integer-valued $M$-tuples $\mathbf{a}=\left[
a_{m};m\in \mathcal{Z}_{M}\right] $ and $\mathbf{b}=\left[ b_{m};m\in 
\mathcal{Z}_{M}\right] $. If $1<a_{m}\leq b_{m}$ for all $m\in \mathcal{Z}%
_{M}$, then $f(\mathbf{a}^{t})\geq f(\mathbf{b}^{t})$. Moreover, $f(\mathbf{a%
}^{t})>f(\mathbf{b}^{t})$ if $1<a_{n}<b_{n}$ for some $n\in \mathcal{Z}_{M}$
and $1<a_{m}\leq b_{m}$ for all the other $m\in \mathcal{Z}_{M}-\{n\}$.

From \emph{Lemma 1}, the smallest two prime factors should be merged to
compose a proper level-$(\Omega (N)-1)$ factorization $N=\prod%
\nolimits_{m=0}^{\Omega (N)-2}A_{m}^{(1)}$ with $A_{0}^{(1)}=P_{0}P_{1}$ and 
$A_{m}^{(1)}=P_{m+1}$ for $m\in \mathcal{Z}_{\Omega (N)-2}^{+}$. This proper
factorization results in the largest family size $\Psi ^{(\text{dpma,}%
1)}(N)=(P_{0}P_{1}-1)\prod\nolimits_{m=2}^{\Omega (N)-1}(P_{m}-1)$. The
corresponding family $\mathcal{G}_{\text{max,}\widetilde{I}}^{(\text{dpma,}%
1)}$ can provide mutually orthogonal order-$\widetilde{I}$ CA sequences with 
$\widetilde{I}\geq \Omega (N)-1$ under \emph{Condition A }and $\widetilde{I}%
\geq \left\lfloor (\Omega (N)-1)/2\right\rfloor $ under \emph{Condition B}.

\emph{B) Proper Level-}$(\Omega (N)-2)$\emph{\ Factorization for }$\Omega
(N)>3$\emph{: }With $N=\prod\nolimits_{m=0}^{\Omega (N)-1}P_{m}$ and $%
\Omega (N)>3$, there are two mutually exclusive methods to factorize $%
N=\prod\nolimits_{m=0}^{\Omega (N)-3}A_{m}$ in order to obtain a level-$%
(\Omega (N)-2)$ factorization. \emph{Method 1} is to choose any three prime
factors from $\{P_{m};m\in \mathcal{Z}_{\Omega (N)}\}$ and merge them into
one composite factor. \emph{Method 2} is to choose any four prime factors
from $\{P_{m};m\in \mathcal{Z}_{\Omega (N)}\}$ and merge them into two
composite factors in pairs. Both methods are detailed below.

\emph{Method 1:} From \emph{Lemma 1}, the smallest three prime factors
should be merged in order to maximize the family size when a level-$(\Omega
(N)-2)$ factorization is obtained by merging three prime factors. This
results in the family size $(P_{0}P_{1}P_{2}-1)\prod\nolimits_{m=3}^{\Omega
(N)-1}(P_{m}-1)$. Thus, one candidate family for $\mathcal{G}_{\text{max,}%
\widetilde{I}}^{(\text{dpma,}2)}$ is based on the candidate factorization $%
A_{0}=P_{0}P_{1}P_{2}$ and $A_{m}=P_{m+2}$ for $m\in \mathcal{Z}_{\Omega
(N)-3}^{+}$.

\emph{Method 2:} When a level-$(\Omega (N)-2)$ factorization is obtained by
merging four prime factors in pairs, the family size can be maximized by
choosing and pairing four prime factors properly based on \emph{Lemma 2}, as
proven in \emph{Appendix B}.

\emph{Lemma 2}: Consider four integers $P_{a}$, $P_{b}$, $P_{c}$ and $P_{d}$%
. If $1<P_{a}\leq P_{b}\leq P_{c}\leq P_{d}$, then $f([P_{a},P_{d}])\times $ 
$f([P_{b},P_{c}])\geq f([P_{a},P_{c}])\times $ $f([P_{b},P_{d}])\geq
f([P_{a},P_{b}])\times $ $f([P_{c},P_{d}])$.

From \emph{Lemma 1} and \emph{Lemma 2}, the smallest four prime factors
should be merged in pairs to form the candidate factorization $%
A_{0}=P_{0}P_{3}$, $A_{1}=P_{1}P_{2}$ and $A_{m+1}=P_{m+3}$ for $m\in 
\mathcal{Z}_{\Omega (N)-4}^{+}$, in order to maximize the family size when a
level-$(\Omega (N)-2)$ factorization is obtained by merging four prime
factors in pairs. Such factorization results in the other candidate family
for $\mathcal{G}_{\text{max,}\widetilde{I}}^{(\text{dpma,}2)}$ having the
family size $(P_{0}P_{3}-1)(P_{1}P_{2}-1)\prod\nolimits_{m=4}^{\Omega
(N)-1}(P_{m}-1)$.

Exclusively, \emph{Method 1} and \emph{Method 2} give two candidate level-$%
(\Omega (N)-2)$ factorizations offering the family sizes $\Psi (N)\times
f([P_{0},P_{1},P_{2}])$ and $\Psi (N)\times f([P_{0},P_{3}])\times
f([P_{1},P_{2}])$, respectively. The factorization that yields the largest
family size is thus a proper level-$(\Omega (N)-2)$ factorization, and can
be adopted to construct family $\mathcal{G}_{\text{max,}\widetilde{I}}^{(%
\text{dpma,}2)}$. \emph{Lemma 3} is proven in \emph{Appendix C} to support
such proper factorization.

\emph{Lemma 3}: Consider four integers $P_{a}$, $P_{b}$, $P_{c}$, $P_{d}$
with $1<P_{a}\leq P_{b}\leq P_{c}\leq P_{d}$. If $P_{b}P_{c}\leq P_{d}$,
then $f([P_{a},P_{b},P_{c}])\geq f([P_{a},P_{d}])\times f([P_{b},P_{c}])$.

From \emph{Lemma 3}, a proper level-$(\Omega (N)-2)$ factorization for $%
N=\prod\nolimits_{m=0}^{\Omega (N)-3}A_{m}^{(2)}$ is obtained by setting $%
A_{0}^{(2)}=P_{0}P_{1}P_{2}$ and $A_{m}^{(2)}=P_{m+2}$ for $m\in \mathcal{Z}%
_{\Omega (N)-3}^{+}$ (i.e., \emph{Method 1}) if $P_{1}P_{2}<P_{3}$, and by
setting $A_{0}^{(2)}=P_{0}P_{3}$, $A_{1}^{(2)}=P_{1}P_{2}$ and $%
A_{m+1}^{(2)}=P_{m+3}$ for $m\in \mathcal{Z}_{\Omega (N)-4}^{+}$ (i.e., 
\emph{Method 2}) otherwise. This proper factorization results in the largest
family size $\Psi ^{(\text{dpma,}2)}(N)=\max
\{(P_{0}P_{1}P_{2}-1)\prod\nolimits_{m=3}^{\Omega
(N)-1}(P_{m}-1),(P_{0}P_{3}-1)(P_{1}P_{2}-1)\prod\nolimits_{m=4}^{\Omega
(N)-1}(P_{m}-1)\}$. Based on the proper level-$(\Omega (N)-2)$
factorization, family $\mathcal{G}_{\text{max,}\widetilde{I}}^{(\text{dpma,}%
2)}$ is constructed by the PMA method and provides mutually orthogonal order-%
$\widetilde{I}$ CA sequences with $\widetilde{I}\geq \Omega (N)-2$ under 
\emph{Condition A }and $\widetilde{I}\geq \left\lfloor (\Omega
(N)-2)/2\right\rfloor $ under \emph{Condition B}.

\emph{C) Exclusively Searching a Proper Level-}$(\Omega (N)-\kappa )$\emph{\
Factorization:} For $\kappa \in \{3,4,...,\Omega (N)-2\}$, it is difficult
to find proper level-$(\Omega (N)-\kappa )$ factorizations in closed-form
expressions. An exclusive search procedure is proposed instead to find such
proper factorizations.

To obtain a proper level-$(\Omega (N)-\kappa )$ factorization, we need to
(i) find all possible factor sets $\{A_{m};m\in \mathcal{Z}_{\Omega
(N)-\kappa }\}$ satisfying $\prod\nolimits_{m=0}^{\Omega (N)-\kappa
-1}A_{m}=\prod\nolimits_{m=0}^{\Omega (N)-1}P_{m}$ by partitioning the
prime factor set $\{P_{m};m\in \mathcal{Z}_{\Omega (N)}\}$ into $\Omega
(N)-\kappa $ groups first and then taking all group products in the
exclusive manner, and (ii) search for a proper factor set $\{A_{m}^{(\kappa
)};m\in \mathcal{Z}_{\Omega (N)-\kappa }\}$ which yields the largest family
size $\Psi ^{(\text{dpma,}\kappa )}(N)=\prod\nolimits_{m=0}^{\Omega
(N)-\kappa -1}(A_{m}^{(\kappa )}-1)$ among all factor sets. In each
partitioning, we denote $\omega _{m}$ as the number of prime factors in the $%
m$-th group, i.e., $\omega _{m}=\Omega (A_{m})$ for $m\in \mathcal{Z}%
_{\Omega (N)-\kappa }$. Thus, each factor set $\{A_{m};m\in \mathcal{Z}%
_{\Omega (N)-\kappa }\}$ is characterized by the corresponding omega pattern 
$\bm{\omega }\triangleq \lbrack \omega _{m};m\in \mathcal{Z}_{\Omega
(N)-\kappa }]$ with $\sum\nolimits_{m=0}^{\Omega (N)-\kappa -1}\omega
_{m}=\Omega (N)$. The exclusive partitioning can be conducted by searching
for all possible omega patterns first and then finding all possible
groupings for each pattern $\bm{\omega }$. To avoid repetitive search, $%
\bm{\omega }$ is limited to have descending entries $\omega _{0}\geq
\omega _{1}\geq ...\geq \omega _{\Omega (N)-\kappa -1}$ in the exclusive
partitioning. In the following, an exclusive search procedure is proposed
accordingly to find a proper level-$(\Omega (N)-\kappa )$ factorization.

\emph{Step 1}: Obtain and store all admissible patterns for $\bm{\omega }
$ under the constraints $\sum\nolimits_{m=0}^{\Omega (N)-\kappa -1}\omega
_{m}=\Omega (N)$ and $\omega _{0}\geq \omega _{1}\geq ...\geq \omega
_{\Omega (N)-\kappa -1}\geq 1$ by the process of integer partitioning in 
\cite[Section 1.1]{Omega Pattern}-\cite{Omega Pattern Alg}.

\emph{Step 2}: Transform the prime factor set $\{P_{m};m\in \mathcal{Z}%
_{\Omega (N)}\}$ into all possible factor sets $\{A_{m};m\in \mathcal{Z}%
_{\Omega (N)-\kappa }\}$ characterized by each admissible pattern $\bm{%
\omega }$ exclusively from Gosper's Hack algorithm \cite[Section 7.1.3]%
{Gosper's Hack}-\cite{Gospe's Hack Alg}. Compute family sizes $%
\prod\nolimits_{m=0}^{\Omega (N)-\kappa -1}(A_{m}-1)$ for all sought factor
sets $\{A_{m};m\in \mathcal{Z}_{\Omega (N)-\kappa }\}$. Store one candidate
factor set which provides the largest family size among all sought factor
sets characterized by each admissible pattern $\bm{\omega }$.

\emph{Step 3}: Find a proper level-$(\Omega (N)-\kappa )$ factor set $%
\{A_{m}^{(\kappa )};m\in \mathcal{Z}_{\Omega (N)-\kappa }\}$ by choosing one
candidate factor set which yields the largest family size $\Psi ^{(\text{%
dpma,}\kappa )}(N)=\prod\nolimits_{m=0}^{\Omega (N)-\kappa
-1}(A_{m}^{(\kappa )}-1)$ among all stored factor sets in \emph{Step 2}. \ \
\ \ \ \ \ \ \ \ \ \ \ \ \ \ \ \ \ \ \ \ \ \ \ \ \ \ \ \ \ \ \ \ \ \ \ \ \ \ \ \ \ \ \ \ \ \ \ \ \ \ \ \ \ \ \ \ \ \ \ \ \ \ $%
\blacksquare $

In \emph{Step 1}, the process of integer partitioning in \cite[Section 1.1]%
{Omega Pattern}-\cite{Omega Pattern Alg} finds all possible patterns for $%
\bm{\omega }$ by dividing the all-one $\Omega (N)$-tuple $%
[1,1,...,1]^{t} $ into the admissible $(\Omega (N)-\kappa )$-tuple $\bm{%
\omega }$ in the exclusive manner. For example, the process finds $%
[3,1,1]^{t}$ and $[2,2,1]^{t}$ by dividing $[1,1,1,1,1]^{t}$ into $[\omega
_{0},\omega _{1},\omega _{2}]^{t}$ for $\Omega (N)=5$ and $\kappa =2$. In 
\emph{Step 2}, Gosper's Hack algorithm transforms an omega pattern to all
possible binary codewords without repetition in the bitwise manner \cite[%
Algorithm 3.1]{Gospe's Hack Alg}, as detailed in \emph{Appendix D}. In \emph{%
Step 3}, a proper level-$(\Omega (N)-\kappa )$ factor set $\{A_{m}^{(\kappa
)};m\in \mathcal{Z}_{\Omega (N)-\kappa }\}$ is found from all stored
candidate factor sets stored in \emph{Step 2} by identifying the largest
family size. This completes the exclusive search procedure.

\emph{D) Near-Proper Level-}$(\Omega (N)-\kappa )$\emph{\ Factorization for }%
$\Omega (N)>4$ \emph{and }$\kappa \in \{3,4,...,\Omega (N)-2\}$\emph{: }A 
\emph{near-proper} level-$(\Omega (N)-\kappa )$ factorization $%
N=\prod\nolimits_{m=0}^{\Omega (N)-\kappa -1}\widetilde{A}_{m}^{(\kappa )}$
with $\widetilde{A}_{0}^{(\kappa )}\leq \widetilde{A}_{1}^{(\kappa )}\leq
...\leq \widetilde{A}_{\Omega (N)-\kappa -1}^{(\kappa )}$ for all $\kappa
\in \mathcal{Z}_{\Omega (N)-2}^{+}-\mathcal{Z}_{2}^{+}$ is proposed here to
construct another degenerate PMA sequence family $\widetilde{\mathcal{G}}_{%
\text{max,}\widetilde{I}}^{(\text{dpma,}\kappa )}$ based on the construction
method of \emph{Proper Level-}$(\Omega (N)-2)$\emph{\ Factorization} in
Subsection III.\emph{B}. Under such near-proper factorization, family $%
\widetilde{\mathcal{G}}_{\text{max,}\widetilde{I}}^{(\text{dpma,}\kappa )}$
exhibits the family size $\widetilde{\Psi }^{(\text{dpma,}\kappa
)}(N)\triangleq \prod\nolimits_{m=0}^{\Omega (N)-\kappa -1}(\widetilde{A}%
_{m}^{(\kappa )}-1)$. Despite $\widetilde{\Psi }^{(\text{dpma,}\kappa
)}(N)\leq \Psi ^{(\text{dpma,}\kappa )}(N)$, the near-proper level-$(\Omega
(N)-\kappa )$ factorization can be obtained simply in a closed-form
expression without resort to exclusive searching.

Following Subsection III.\emph{B}, a \emph{near-proper} level-$(\Omega
(N)-\kappa )$ factorization for $N=\prod\nolimits_{m=0}^{\Omega (N)-\kappa
-1}\widetilde{A}_{m}^{(\kappa )}$ is obtained from a given level-$(\Omega
(N)-\kappa +2)$ factorization $N=\prod\nolimits_{m=0}^{\Omega (N)-\kappa +1}%
\widetilde{A}_{m}^{(\kappa -2)}$ with the arranged order $\widetilde{A}%
_{0}^{(\kappa -2)}\leq \widetilde{A}_{1}^{(\kappa -2)}\leq ...\leq 
\widetilde{A}_{\Omega (N)-\kappa +1}^{(\kappa -2)}$. Specifically, a
near-proper level-$(\Omega (N)-\kappa )$ factorization for $%
N=\prod\nolimits_{m=0}^{\Omega (N)-\kappa -1}\widetilde{A}_{m}^{(\kappa )}$
is obtained by setting $\widetilde{A}_{0}^{(\kappa )}=\widetilde{A}%
_{0}^{(\kappa -2)}\widetilde{A}_{1}^{(\kappa -2)}\widetilde{A}_{2}^{(\kappa
-2)}$ and $\widetilde{A}_{m}^{(\kappa )}=\widetilde{A}_{m+2}^{(\kappa -2)}$
for $m\in \mathcal{Z}_{\Omega (N)-\kappa -1}^{+}$ if $\widetilde{A}%
_{1}^{(\kappa -2)}\widetilde{A}_{2}^{(\kappa -2)}<\widetilde{A}_{3}^{(\kappa
-2)}$, and by setting $\widetilde{A}_{0}^{(\kappa )}=\widetilde{A}%
_{0}^{(\kappa -2)}\widetilde{A}_{3}^{(\kappa -2)}$, $\widetilde{A}%
_{1}^{(\kappa )}=\widetilde{A}_{1}^{(\kappa -2)}\widetilde{A}_{2}^{(\kappa
-2)}$ and $\widetilde{A}_{m+1}^{(\kappa )}=\widetilde{A}_{m+3}^{(\kappa -2)}$
for $m\in \mathcal{Z}_{\Omega (N)-\kappa -2}^{+}$ otherwise. For $\kappa \in 
\mathcal{Z}_{2}^{+}$, $\{\widetilde{A}_{m}^{(\kappa )};m\in \mathcal{Z}%
_{\Omega (N)-\kappa }\}$ is initially assigned by $\widetilde{A}%
_{m}^{(\kappa )}=A_{m}^{(\kappa )}$ where $N=\prod\nolimits_{m=0}^{\Omega
(N)-\kappa -1}A_{m}^{(\kappa )}$ is a proper level-$(\Omega (N)-\kappa )$
factorization with the arranged order $A_{0}^{(\kappa )}\leq A_{1}^{(\kappa
)}\leq ...\leq A_{\Omega (N)-\kappa -1}^{(\kappa )}$.

Notably, families $\widetilde{\mathcal{G}}_{\text{max,}\widetilde{I}}^{(%
\text{dpma,}\kappa )}$ contain $\widetilde{\Psi }^{(\text{dpma,}\kappa
)}(N)/(\widetilde{A}_{\max }^{(\kappa )}-1)$ mutually exclusive CS sequence
subfamilies and each subfamily contains $\widetilde{A}_{\max }^{(\kappa )}-1$%
\ sequences generated by cyclically shifting the inverse DFT of a subfamily
leader with family CSD $\varpi _{\widetilde{\mathcal{G}}}^{(\kappa
)}\triangleq N/\widetilde{A}_{\max }^{(\kappa )}$, where $\widetilde{A}%
_{\max }^{(\kappa )}\triangleq \max_{m\in \mathcal{Z}_{\Omega (N)-\kappa }}%
\widetilde{A}_{m}^{(\kappa )}$.

\emph{E) Some Examples of Families }$\mathcal{G}_{\text{max,}\widetilde{I}%
}^{(\text{dpma,}\kappa )}$\emph{\ and }$\widetilde{\mathcal{G}}_{\text{max,}%
\widetilde{I}}^{(\text{dpma,}\kappa )}$\emph{:} Based on proper level-$%
(\Omega (N)-\kappa )$ factorization $N=$ $\prod\nolimits_{m=0}^{\Omega
(N)-\kappa -1}A_{m}^{(\kappa )}$ and near-proper level-$(\Omega (N)-\kappa )$
factorization $N=\prod\nolimits_{m=0}^{\Omega (N)-\kappa -1}\widetilde{A}%
_{m}^{(\kappa )}$, family $\mathcal{G}_{\text{max,}\widetilde{I}}^{(\text{%
dpma,}\kappa )}$ and family $\widetilde{\mathcal{G}}_{\text{max,}\widetilde{I%
}}^{(\text{dpma,}\kappa )}$ can be constructed to contain orthogonal order-$%
\widetilde{I}$ CA sequences with family sizes $\prod\nolimits_{m=0}^{\Omega
(N)-\kappa -1}(A_{m}^{(\kappa )}-1)$ and $\prod\nolimits_{m=0}^{\Omega
(N)-\kappa -1}(\widetilde{A}_{m}^{(\kappa )}-1)$, respectively. Since there
is only one level-$\Omega (N)$ factorization, family $\mathcal{G}_{I}^{(%
\text{pma})}$ is essentially constructed under a proper level-$(\Omega
(N)-\kappa )$ factorization with $\kappa =0$. The developed degenerate PMA
families $\mathcal{G}_{\text{max,}\widetilde{I}}^{(\text{dpma,}\kappa )}$%
\emph{\ }and\emph{\ }$\widetilde{\mathcal{G}}_{\text{max,}\widetilde{I}}^{(%
\text{dpma,}\kappa )}$ are listed in Table I for example sequence lengths
adopted by the MIMO channel sounding application in 5G-NR \cite[Section
6.4.1.4.3]{5G}, where the achieved family size, the supporting factor set,
and the family CSD are demonstrated for each family. As shown, $\widetilde{%
\mathcal{G}}_{\text{max,}\widetilde{I}}^{(\text{dpma,}\kappa )}$ and $%
\mathcal{G}_{\text{max,}\widetilde{I}}^{(\text{dpma,}\kappa )}$ provide much
larger family sizes than $\mathcal{G}_{I}^{(\text{pma})}$ and offer the
larger family sizes as $\kappa $ increases, but they may entail reduced SD
order $\widetilde{I}\geq \Omega (N)-\kappa $ under \emph{Condition A} and $%
\widetilde{I}\geq \left\lfloor (\Omega (N)-\kappa )/2\right\rfloor $ under 
\emph{Condition B}. For a fixed $\kappa \in \{3,4,...,\Omega (N)-2\}$, the
family size of $\widetilde{\mathcal{G}}_{\text{max,}\widetilde{I}}^{(\text{%
dpma,}\kappa )}$ is the same as or very close to the family size of $%
\mathcal{G}_{\text{max,}\widetilde{I}}^{(\text{dpma,}\kappa )}$. The latter
reveals the advantage of near-proper level-$(\Omega (N)-\kappa )$
factorization for $\kappa \in \{3,4,...,\Omega (N)-2\}$ in that the
closed-form expressions are available for factorization.

For the channel sounding application in 5G NR, \emph{nonorthogonal}
subfamilies of cyclically-shiftable ZC sequences generated from different
root indices are adopted to enable simultaneous MIMO channel estimation in
multiple cells/sectors environments. All cyclically-shiftable ZC sequences
in a subfamily are generated by cyclically shifting the inverse DFT of a
single-root ZC sequence under the restriction that the minimum CSD $\varpi _{%
\text{min}}$ is guaranteed for every shift to avoid sequence identification
ambiguity \cite[Section 6.4.1.4.1]{5G}. However, family $\mathcal{G}_{\text{%
max,}\widetilde{I}}^{(\text{dpma,}\kappa )}$ contains $\Psi ^{(\text{dpma,}%
\kappa )}(N)/(A_{\max }^{(\kappa )}-1)$ \emph{orthogonal} CS sequence
subfamilies and each subfamily contains $A_{\max }^{(\kappa )}-1$\
cyclically-shiftable sequences generated by cyclically shifting the inverse
DFT of a subfamily leader with family CSD $\varpi _{\mathcal{G}}^{(\kappa
)}=N/A_{\max }^{(\kappa )}$. Notably, not every sequence in family $\mathcal{%
G}_{\text{max,}\widetilde{I}}^{(\text{dpma,}\kappa )}$ can be adopted if $%
\varpi _{\mathcal{G}}^{(\kappa )}$ is smaller than $\varpi _{\text{min}}$
required by the channel sounding application. The same concern exists with
family $\widetilde{\mathcal{G}}_{\text{max,}\widetilde{I}}^{(\text{dpma,}%
\kappa )}$. For example, family $\mathcal{G}_{\text{max,}\widetilde{I}}^{(%
\text{dpma,}5)}$ in Table I(c) contains $\Psi ^{(\text{dpma,}%
5)}(288)/(A_{\max }^{(5)}-1)=15$ CS sequence subfamilies and each subfamily
contains $\widetilde{A}_{\max }^{(5)}-1=17$ sequences with $\varpi _{%
\mathcal{G}}^{(5)}=16$. Due to $\varpi _{\text{min}}=24$, at most $%
\left\lfloor (A_{\max }^{(5)}-1)/\left\lceil \varpi _{\text{min}}/\varpi _{%
\mathcal{G}}^{(5)}\right\rceil \right\rfloor =8$ sequences can be adopted in
each subfamily to satisfy the CSD restriction and accordingly at most $120$
orthogonal sequences from family $\mathcal{G}_{\text{max,}\widetilde{I}}^{(%
\text{dpma,}5)}$ can be used as the sounding sequences under the restriction 
$\varpi _{\text{min}}=24$. As such, the number $\Psi (N|\varpi _{\text{min}})
$ is also given for each family in Table I to show the maximum number of
sounding sequences available for use in the channel sounding application
under the CSD restriction by $\varpi _{\text{min}}$. Since all sequences are
mutually orthogonal, these degenerate PMA sequence families are useful to
mitigate the effect of pilot contamination in simultaneous MIMO channel
estimation (e.g., \cite{MIMO CS 1}).

\begin{table}\centering%
\caption{The family sizes and the numbers of available sounding sequences provided by families  $\mathcal{G}_{\rm{max},\widetilde{\emph{I}}}^{(\rm{dpma},\kappa)}$ and 
$\widetilde{\mathcal{G}}_{\rm{max},\widetilde{\emph{I}}}^{(\rm{dpma},\kappa)}$ for sequence lengths 
($\text{a}$) $N=48$, ($\text{b}$) $N=144$, and ($\text{c}$) $N=288$ adopted for channel sounding  application in 5G NR standard. 
The associated factor sets $\{A_{m}^{(\kappa)};m\in\mathcal{Z}_{\Omega(N)-\kappa}\}$ and
$\{\widetilde{A}_{m}^{(\kappa)};m\in\mathcal{Z}_{\Omega(N)-\kappa}\}$ are also demonstrated.}

\begin{tabular}{c}
(a) $N=48$ $(\Omega (N)=5,\varpi _{\text{min}}=4)$ \\ 
\begin{tabular}{c||c|c}
\hline
\begin{tabular}{c}
Family \\ 
(Family CSD)%
\end{tabular}
& 
\begin{tabular}{c}
Family Size \\ 
($\Psi (N|\varpi _{\text{min}})$)%
\end{tabular}
& Factor Set \\ \hline\hline
$\mathcal{G}_{I}^{(\text{pma})}$ ($16$) & $2$ ($2$) & $\{2,2,2,2,3\}$ \\ 
\hline
$\mathcal{G}_{\text{max,}\widetilde{I}}^{(\text{dpma,}1)}$ ($12$) & $6$ ($6$)
& $\{2,2,3,4\}$ \\ \hline
$\mathcal{G}_{\text{max,}\widetilde{I}}^{(\text{dpma,}2)}$ ($12$) & $18$ ($%
18 $) & $\{3,4,4\}$ \\ \hline
\begin{tabular}{c}
$\mathcal{G}_{\text{max,}\widetilde{I}}^{(\text{dpma,}3)},\widetilde{%
\mathcal{G}}_{\text{max,}\widetilde{I}}^{(\text{dpma,}3)}$ \\ 
($6,6$)%
\end{tabular}
& $35,35$ ($35,35$) & $\{6,8\},\{6,8\}$ \\ \hline
\end{tabular}%
\end{tabular}

\begin{tabular}{c}
(b) $N=144$ $(\Omega (N)=6,\varpi _{\text{min}}=12)$ \\ 
\begin{tabular}{c||c|c}
\hline
\begin{tabular}{c}
Family \\ 
(Family CSD)%
\end{tabular}
& 
\begin{tabular}{c}
Family Size \\ 
($\Psi (N|\varpi _{\text{min}})$)%
\end{tabular}
& Factor Set \\ \hline\hline
$\mathcal{G}_{I}^{(\text{pma})}$ ($48$) & $4$ ($4$) & $\{2,2,2,2,3,3\}$ \\ 
\hline
$\mathcal{G}_{\text{max,}\widetilde{I}}^{(\text{dpma,}1)}$ ($36$) & $12$ ($%
12 $) & $\{2,2,3,3,4\}$ \\ \hline
$\mathcal{G}_{\text{max,}\widetilde{I}}^{(\text{dpma,}2)}$ ($36$) & $36$ ($%
36 $) & $\{3,3,4,4\}$ \\ \hline
\begin{tabular}{c}
$\mathcal{G}_{\text{max,}\widetilde{I}}^{(\text{dpma,}3)},\widetilde{%
\mathcal{G}}_{\text{max,}\widetilde{I}}^{(\text{dpma,}3)}$ \\ 
($24,24$)%
\end{tabular}
& 
\begin{tabular}{c}
$75,75$ \\ 
($75,75$)%
\end{tabular}
& $\{4,6,6\},\{4,6,6\}$ \\ \hline
\begin{tabular}{c}
$\mathcal{G}_{\text{max,}\widetilde{I}}^{(\text{dpma,}4)},\widetilde{%
\mathcal{G}}_{\text{max,}\widetilde{I}}^{(\text{dpma,}4)}$ \\ 
($12,12$)%
\end{tabular}
& 
\begin{tabular}{c}
$121,121$ \\ 
($121,121$)%
\end{tabular}
& $\{12,12\},\{12,12\}$ \\ \hline
\end{tabular}%
\end{tabular}

\begin{tabular}{c}
(c) $N=288$ $(\Omega (N)=7,\varpi _{\text{min}}=24)$ \\ 
\begin{tabular}{c||c|c}
\hline
\begin{tabular}{c}
Family \\ 
(Family CSD)%
\end{tabular}
& 
\begin{tabular}{c}
Family Size \\ 
($\Psi (N|\varpi _{\text{min}})$)%
\end{tabular}
& Factor Set \\ \hline\hline
$\mathcal{G}_{I}^{(\text{pma})}$ ($96$) & $4$ ($4$) & $\{2,2,2,2,2,3,3\}$ \\ 
\hline
$\mathcal{G}_{\text{max,}\widetilde{I}}^{(\text{dpma,}1)}$ ($72$) & $12$ ($%
12 $) & $\{2,2,2,3,3,4\}$ \\ \hline
$\mathcal{G}_{\text{max,}\widetilde{I}}^{(\text{dpma,}2)}$ ($72$) & $36$ ($%
36 $) & $\{2,3,3,4,4\}$ \\ \hline
\begin{tabular}{c}
$\mathcal{G}_{\text{max,}\widetilde{I}}^{(\text{dpma,}3)},\widetilde{%
\mathcal{G}}_{\text{max,}\widetilde{I}}^{(\text{dpma,}3)}$ \\ 
($48,48$)%
\end{tabular}
& 
\begin{tabular}{c}
$90,90$ \\ 
($90,90$)%
\end{tabular}
& $%
\begin{array}{l}
\{3,4,4,6\}, \\ 
\{3,4,4,6\}%
\end{array}%
$ \\ \hline
\begin{tabular}{c}
$\mathcal{G}_{\text{max,}\widetilde{I}}^{(\text{dpma,}4)},\widetilde{%
\mathcal{G}}_{\text{max,}\widetilde{I}}^{(\text{dpma,}4)}$ \\ 
($36,32$)%
\end{tabular}
& 
\begin{tabular}{c}
$175,168$ \\ 
($175,168$)%
\end{tabular}
& $\{6,6,8\},\{4,8,9\}$ \\ \hline
\begin{tabular}{c}
$\mathcal{G}_{\text{max,}\widetilde{I}}^{(\text{dpma,}5)},\widetilde{%
\mathcal{G}}_{\text{max,}\widetilde{I}}^{(\text{dpma,}5)}$ \\ 
($16,16$)%
\end{tabular}
& 
\begin{tabular}{c}
$255,255$ \\ 
($120,120$)%
\end{tabular}
& $\{16,18\},\{16,18\}$ \\ \hline
\end{tabular}%
\end{tabular}%
\end{table}%

\section{Families $\protect\widehat{\mathcal{G}}_{I}^{(\rm{apma})}$, $%
\protect\widehat{\mathcal{G}}_{\rm{max,}\protect\widetilde{\emph{I}}}^{(\rm{%
dpma,}\protect\kappa )}$, and $\protect\widehat{\mathcal{G}}_{\rm{max,}%
\protect\widetilde{\emph{I}}}^{(\rm{adpma,}\protect\kappa )}$}

In this section, we consider the sequence length $N$ meeting $\widetilde{%
\Omega }(N)>\Omega (N)$ and construct new families $\widehat{\mathcal{G}}_{%
\text{max,}\widetilde{I}}^{(\text{dpma,}\kappa )}$and $\widehat{\mathcal{G}}%
_{\text{max,}\widetilde{I}}^{(\text{adpma,}\kappa )}$ for $\kappa \in 
\mathcal{Z}_{\widetilde{\Omega }(N)-1}^{+}$. Notably, any orthogonal
sequence in a new family $\widehat{\mathcal{G}}_{\text{max,}\widetilde{I}}^{(%
\text{dpma,}\kappa )}$or $\widehat{\mathcal{G}}_{\text{max,}\widetilde{I}}^{(%
\text{adpma,}\kappa )}$ can not be obtained from cyclically shifting the
inverse DFT of another sequence in the same family, due to the proper
decomposition of $N$.

\emph{A) Degenerate PMA Sequence Family }$\widehat{\mathcal{G}}_{\text{max,}%
\widetilde{I}}^{(\text{dpma,}\kappa )}$ \emph{for} $\kappa \in \mathcal{Z}_{%
\widetilde{\Omega }(N)-1}^{+}$\emph{: }With a given $\kappa \in \mathcal{Z}_{%
\widetilde{\Omega }(N)-1}^{+}$, the degenerate PMA sequence family $\widehat{%
\mathcal{G}}_{\text{max,}\widetilde{I}}^{(\text{dpma,}\kappa )}$ can be
constructed by virtue of the \emph{combined proper} level-$(\widetilde{%
\Omega }(N)-\kappa )$ factorization for sequence $\widehat{\mathcal{G}}_{I}$%
, which is composed of \emph{individual proper} level-$(\Omega (\widetilde{N}%
^{(\rho )})-\kappa )$ factorizations for all component subsequences of
lengths $\widetilde{N}^{(0)}$, $\widetilde{N}^{(1)}$,..., $\widetilde{N}%
^{(L-1)}$ under the proper decomposition $N=\sum_{\rho \in \mathcal{Z}_{L}}%
\widetilde{N}^{(\rho )}$ yielding $\widetilde{\Omega }(N)$ (see (\ref{MPO}))
with $\widetilde{N}^{(0)}\geq \widetilde{N}^{(1)}\geq ...\geq \widetilde{N}%
^{(L-1)}\geq 2$. For each $\rho \in \mathcal{Z}_{L}$, the proper level-$%
(\Omega (\widetilde{N}^{(\rho )})-\kappa )$ factorization $\widetilde{N}%
^{(\rho )}=\prod\nolimits_{m=0}^{\Omega (\widetilde{N}^{(\rho )})-\kappa
-1}A_{m}^{(\rho ,\kappa )}$ is used to construct family $\mathcal{G}_{\text{%
max,}\widetilde{I}^{\prime }}^{(\text{dpma,}\kappa )}$ with subsequence
length $\widetilde{N}^{(\rho )}$, where the factors $A_{m}^{(\rho ,\kappa )}$
are not all primes. Each orthogonal sequence in family $\widehat{\mathcal{G}}%
_{\text{max,}\widetilde{I}}^{(\text{dpma,}\kappa )}$ is thus composed by a
concatenation of $L$ PMA subsequences $\bm{\chi }_{0}$, $\bm{\chi }%
_{1}$,..., $\bm{\chi }_{L-1}$ of lengths $\widetilde{N}^{(0)}$, $%
\widetilde{N}^{(1)}$,..., $\widetilde{N}^{(L-1)}$, respectively. Such family 
$\widehat{\mathcal{G}}_{\text{max,}\widetilde{I}}^{(\text{dpma,}\kappa )}$
has the family size $\widehat{\Psi }^{(\text{dpma,}\kappa )}(N)\triangleq
\min_{\rho \in \mathcal{Z}_{L}}$ $\prod\nolimits_{m=0}^{\Omega (\widetilde{N%
}^{(\rho )})-\kappa -1}(A_{m}^{(\rho ,\kappa )}-1)$. Notably, $\widehat{%
\mathcal{G}}_{\text{max,}\widetilde{I}}^{(\text{dpma,}\kappa )}$ tends to
offer the larger family size as $\kappa $ is increased, but it may reduce
the SD order $\widetilde{I}\geq \widetilde{\Omega }(N)-\kappa $ under \emph{%
Condition A} and $\widetilde{I}\geq \left\lfloor (\widetilde{\Omega }%
(N)-\kappa )/2\right\rfloor $ under \emph{Condition B}.

\emph{B) Augmented PMA Sequence Families }$\widehat{\mathcal{G}}_{I}^{(\text{%
apma})}$ \emph{and} $\widehat{\mathcal{G}}_{\text{max,}\widetilde{I}}^{(%
\text{adpma,}\kappa )}$ \emph{for} $\kappa \in \mathcal{Z}_{\widetilde{%
\Omega }(N)-1}^{+}$\emph{: }Augmented PMA sequence family $\widehat{\mathcal{%
G}}_{I}^{(\text{apma})}$ expands from family $\widehat{\mathcal{G}}_{I}^{(%
\text{pma})}$ with the same SD order and a double family size, by virtue of
phase-rotating every existing sequence in family $\widehat{\mathcal{G}}%
_{I}^{(\text{pma})}$ to generate more orthogonal sequence members.
Similarly, augmented degenerate PMA sequence family $\widehat{\mathcal{G}}_{%
\text{max,}\widetilde{I}}^{(\text{adpma,}\kappa )}$ expands from $\widehat{%
\mathcal{G}}_{\text{max,}\widetilde{I}}^{(\text{dpma,}\kappa )}$ and offers
a double family size while sustaining the same SD order. In what follows,
the phase-rotating method constructing family $\widehat{\mathcal{G}}_{I}^{(%
\text{apma})}$ is described in detail, and such method is also applied to
construct family $\widehat{\mathcal{G}}_{\text{max,}\widetilde{I}}^{(\text{%
adpma,}\kappa )}$.

Consider one sequence $\widehat{\mathcal{G}}_{I}$ in family $\widehat{%
\mathcal{G}}_{I}^{(\text{pma})}$, which is described by $\bm{\chi }=[%
\bm{\chi }_{0}^{t},\bm{\chi }_{1}^{t},...,\bm{\chi }%
_{L-1}^{t}]^{t}$. From this sequence, one extra order-$I$ CA sequence $%
\widehat{\mathcal{G}}_{I}$ can be obviously constructed by rotating the
phases of the subsequences $\bm{\chi }_{\rho }$ and described by $%
\bm{\chi }^{\bm{\theta }}=[e^{j\theta _{0}}\bm{\chi }%
_{0}^{t},e^{j\theta _{1}}\bm{\chi }_{1}^{t},...,e^{j\theta _{L-1}}%
\bm{\chi }_{L-1}^{t}]^{t}$ where $\bm{\theta }=\left[ \theta _{\rho
};\rho \in \mathcal{Z}_{L}\right] $ is the rotating phase vector. Due to the
PMA construction of $\widehat{\mathcal{G}}_{I}^{(\text{pma})}$, such
phase-rotated sequence is mutually orthogonal to all the other PMA sequences
in family $\widehat{\mathcal{G}}_{I}^{(\text{pma})}$ as well as their
phase-rotated sequences. To ensure that two sequences $\widehat{\mathcal{G}}%
_{I}$ described by $\bm{\chi }$ and $\bm{\chi }^{\bm{\theta }}$
are mutually orthogonal, $\bm{\theta }$ should be chosen to meet the
orthogonality condition $\bm{\chi }^{h}\bm{\chi }^{\bm{\theta }%
}=\sum\nolimits_{\rho \in \mathcal{Z}_{L}}\bm{\chi }_{\rho }^{h}\bm{%
\chi }_{\rho }e^{j\theta _{\rho }}=0$, or equivalently%
\begin{equation}
\sum\limits_{\rho \in \mathcal{Z}_{L}}\widetilde{N}^{(\rho )}e^{j\theta
_{\rho }}=0.  \label{12}
\end{equation}%
Notably, $\sum\nolimits_{\rho \in \mathcal{Z}_{L}}\widetilde{N}^{(\rho
)}e^{j\theta _{\rho }}$ can be regarded as the sum of $L$ vectors $%
\widetilde{N}^{(\rho )}e^{j\theta _{\rho }}$ in the complex plane. As
indicated by (\ref{12}), these $L$ vectors should be connected to form an $L$%
-edge cyclic polygon in the complex plane. According to the cyclic polygon
theorem in \cite[Theorem 1]{Polygon}, a solution to (\ref{12}) exists iff
all edge lengths $\widetilde{N}^{(\rho )}$ meet following restriction.%
\begin{equation*}
\text{\emph{Restriction A: }}L\geq 3\text{ and }\max_{\rho \in \mathcal{Z}%
_{L}}\widetilde{N}^{(\rho )}<\frac{1}{2}\sum\limits_{\rho \in \mathcal{Z}%
_{L}}\widetilde{N}^{(\rho )}.
\end{equation*}%
When \emph{Restriction A} is met, the solution of $\bm{\theta }$ to (\ref%
{12}) can be obtained by invoking the following procedure.

\emph{B.1)} \emph{Procedure to Finding a Proper }$\bm{\theta }$\emph{: }%
This procedure is based on the\ bisection method to constructing a cyclic
polygon given the edge lengths and obtaining the arc angles corresponding to
the given edge lengths simultaneously \cite[Section 1]{Polygon}. Under \emph{%
Restriction A,} there must exist a cyclic polygon with prescribed edge
lengths $\widetilde{N}^{(0)}$, $\widetilde{N}^{(1)}$, ..., $\widetilde{N}%
^{(L-1)}$.\footnote{%
A polygon is said to be cyclic if all vertices of this polygon can
circumscribe a circle.} Denote $\xi $ as the radius of the circumscribed
circle and $\vartheta _{\rho }$ as the arc angle corresponding to the edge
length $\widetilde{N}^{(\rho )}$ of the circumscribed circle. Notably, $\xi $
and $\{\vartheta _{\rho }\}$ are necessary to calculate $\{\theta _{\rho }\}$
\cite[Section 1]{Polygon}. Moreover, we denote $\widehat{\xi }$, $\widehat{%
\vartheta }_{\rho }$, and $\widehat{\theta }_{\rho }$ as the estimates of $%
\xi $, $\vartheta _{\rho }$, and $\theta _{\rho }$, respectively, in order
to approach (\ref{12}). The following procedure is then used to find $\{%
\widehat{\theta }_{\rho }\}$ which can approach (\ref{12}) within a
predetermined accuracy $\epsilon $ \cite[Section 1]{Polygon}.

\emph{Step 0}: Let $\theta _{\text{min}}=0$, $\theta _{\text{max}}=2\pi $,
and $\epsilon $ be a small positive real number close to zero.

\emph{Step 1}: Let $\theta _{\text{mid}}=\frac{1}{2}(\theta _{\text{min}%
}+\theta _{\text{max}})$ and $\widehat{\xi }=\frac{\widetilde{N}^{(0)}}{%
2\sin (\theta _{\text{mid}}/2)}$.

\emph{Step 2}: Let $\widehat{\vartheta }_{0}=\theta _{\text{mid}}$ and $%
\widehat{\vartheta }_{\rho }=\arccos (1-\frac{(\widetilde{N}^{(\rho )})^{2}}{%
2\widehat{\xi }^{2}})$ for all $\rho \in \mathcal{Z}_{L-1}^{+}$.

\emph{Step 3}: There are three cases in this step.

\qquad \emph{Case 1}: If $|\Sigma _{\rho \in \mathcal{Z}_{L}}\widehat{%
\vartheta }_{\rho }-2\pi |\leq \epsilon $, go to \emph{Step 4} directly.

\qquad \emph{Case 2}: If $\Sigma _{\rho \in \mathcal{Z}_{L}}\widehat{%
\vartheta }_{\rho }<2\pi -\epsilon $, let $\theta _{\text{min}}=\theta _{%
\text{mid}}$ and go back to \emph{Step 1}.

\qquad \emph{Case 3}: If $\Sigma _{\rho \in \mathcal{Z}_{L}}\widehat{%
\vartheta }_{\rho }>2\pi +\epsilon $, let $\theta _{\text{max}}=\theta _{%
\text{mid}}$ and go back to \emph{Step 1}.

\emph{Step 4}: Let $\widehat{\theta }_{0}=0$ and $\widehat{\theta }_{\rho }=%
\widehat{\theta }_{\rho -1}+\frac{1}{2}(\widehat{\vartheta }_{\rho -1}+%
\widehat{\vartheta }_{\rho })$ for all $\rho \in \mathcal{Z}_{L-1}^{+}$. \ \
\ \ \ \ \ \ \ \ \ \ \ \ \ \ \ \ \ \ \ \ \ \ \ \ \ \ \ \ \ \ \ \ \ \ \ \ \ \
\ \ \ \ \ \ \ \ \ \ \ \ \ \ \ \ \ $\blacksquare $

When $\Sigma _{\rho \in \mathcal{Z}_{L}}\widehat{\vartheta }_{\rho }<2\pi
-\epsilon $ occurs, all prescribed edge lengths $\widetilde{N}^{(\rho )}$
can not make a cyclic polygon and thus the estimated radius $\widehat{\xi }$
is larger than the actual radius $\xi $. Since $\arccos $ is a monotonically
decreasing function and $\widehat{\xi }>\xi $, $\widehat{\vartheta }_{\rho
}<\vartheta _{\rho }$ for all $\rho \in \mathcal{Z}_{L}$ in this case.
Conversely, $\widehat{\vartheta }_{\rho }>\vartheta _{\rho }$ for all $\rho
\in \mathcal{Z}_{L}$ when $\Sigma _{\rho \in \mathcal{Z}_{L}}\widehat{%
\vartheta }_{\rho }>2\pi +\epsilon $. Through multiple iterations, $\widehat{%
\vartheta }_{\rho }$ can be obtained for all $\rho \in \mathcal{Z}_{L}$. At 
\emph{Step 4}, the absolute error $|\Sigma _{\rho }\vartheta _{\rho }-\Sigma
_{\rho }\widehat{\vartheta }_{\rho }|$ is limited to be within $\epsilon $.
Thus, $\widehat{\theta }_{\rho }$ for all $\rho \in \mathcal{Z}_{L}$ can be
estimated within an accuracy $\epsilon $.

When the proper decomposition $N=\sum_{\rho \in \mathcal{Z}_{L}}\widetilde{N}%
^{(\rho )}$ for a given family $\widehat{\mathcal{G}}_{I}^{(\text{pma})}$
meets \emph{Restriction A}, family $\widehat{\mathcal{G}}_{I}^{(\text{apma}%
)} $ can be constructed by including all orthogonal order-$I$ CA sequences
in family $\widehat{\mathcal{G}}_{I}^{(\text{pma})}$ with size $\widehat{%
\Psi }(N)$ and augmenting another $\widehat{\Psi }(N)$ orthogonal order-$I$
CA sequences, where each augmented sequence described by $\bm{\chi }^{%
\bm{\theta }}$ can be constructed for each sequence $\widehat{\mathcal{G}%
}_{I}$ described by $\bm{\chi }$ with $\bm{\theta }$ obtained by
invoking the aforementioned procedure. Apparently, all augmented sequences
are mutually orthogonal and also orthogonal to all sequences in family $%
\widehat{\mathcal{G}}_{I}^{(\text{pma})}$. This results in the family size $2%
\widehat{\Psi }(N)$ and the same SD order for family $\widehat{\mathcal{G}}%
_{I}^{(\text{apma})}$. Similarly, family $\widehat{\mathcal{G}}_{\text{max,}%
\widetilde{I}}^{(\text{adpma,}\kappa )}$ for $\kappa \in \mathcal{Z}_{%
\widetilde{\Omega }(N)-1}^{+}$ can be augmented from a given family $%
\widehat{\mathcal{G}}_{\text{max,}\widetilde{I}}^{(\text{dpma,}\kappa )}$
and has the family size $2\widehat{\Psi }^{(\text{dpma,}\kappa )}(N)$ under
the same proper decomposition $N=\sum_{\rho \in \mathcal{Z}_{L}}\widetilde{N}%
^{(\rho )}$, while sustaining the same SD order. Notably, $\widehat{\mathcal{%
G}}_{I}^{(\text{apma})}$ can be regarded as a special case of $\widehat{%
\mathcal{G}}_{\text{max,}\widetilde{I}}^{(\text{adpma,}\kappa )}$ with $%
\kappa =0$ and $\widetilde{I}=I$ since $\widehat{\mathcal{G}}_{I}^{(\text{%
apma})}$ is essentially constructed under combined proper level-$\widetilde{%
\Omega }(N)$ factorization for $\widehat{\mathcal{G}}_{\text{max,}I}^{(\text{%
adpma,}0)}$.

\begin{table}\centering%
\caption{The family sizes provided by $\widehat{\mathcal{G}}_{\rm{max},\widetilde{\emph{I}}}^{(\rm{dpma},\kappa)}$ 
and $\widehat{\mathcal{G}}_{\rm{max},\widetilde{\emph{I}}}^{(\rm{adpma},\kappa)}$ with sequence lengths 
$(a) N=139$, $(b) N=571$, $(c) N=839$, and $(d) N=1151$ for RA application in 5G NR. In searching $\bf\theta$, 
the accuracy measure is set to $\epsilon = 10^{-9}$. The proper factor sets for all subsequence lengths $\widetilde{N}^{(0)}$, 
$\widetilde{N}^{(1)}$, ..., $\widetilde{N}^{(L-1)}$ and the sought $\widehat{\bf\theta}$ are also demonstrated. The largest achievable family size $\Psi_{\text{max}}(N)$ 
under $Condition$ $B$ is given for benchmarking.}

\begin{tabular}{c}
\begin{tabular}{c}
(a) $N=139$ with $(\widetilde{N}^{(0)},\widetilde{N}^{(1)},\widetilde{N}%
^{(2)})=(50,45,44)$, \\ 
$\widetilde{\Omega }(N)=3$, and $\{\widehat{\theta }_{0},\widehat{\theta }%
_{1},\widehat{\theta }_{2}\}=\{0^{\circ },125.12^{\circ },236.77^{\circ }\}$%
\end{tabular}
\\ 
\begin{tabular}{c||c|c}
\hline
Family & 
\begin{tabular}{c}
Family Size \\ 
($\Psi _{\text{max}}(N)$)%
\end{tabular}
& 
\begin{tabular}{c}
Factor Sets of $\widetilde{N}^{(0)},$ \\ 
$\widetilde{N}^{(1)},$ and $\widetilde{N}^{(2)}$%
\end{tabular}
\\ \hline\hline
$\widehat{\mathcal{G}}_{I}^{(\text{pma})},\widehat{\mathcal{G}}_{I}^{(\text{%
apma})}$ & 
\begin{tabular}{c}
$10,20$ \\ 
($137$)%
\end{tabular}
& 
\begin{tabular}{c}
$\{2,5,5\},\{3,3,$ \\ 
$5\},\{2,2,11\}$%
\end{tabular}
\\ \hline
$\widehat{\mathcal{G}}_{\text{max,}\widetilde{I}}^{(\text{dpma,}1)},\widehat{%
\mathcal{G}}_{\text{max,}\widetilde{I}}^{(\text{adpma,}1)}$ & 
\begin{tabular}{c}
$30,60$ \\ 
($137$)%
\end{tabular}
& 
\begin{tabular}{c}
$\{5,10\},$ \\ 
$\{5,9\},\{4,11\}$%
\end{tabular}
\\ \hline
\end{tabular}%
\end{tabular}

\begin{tabular}{c}
\begin{tabular}{c}
(b) $N=571$ with $(\widetilde{N}^{(0)},\widetilde{N}^{(1)},\widetilde{N}%
^{(2)})=(225,196,150)$, \\ 
$\widetilde{\Omega }(N)=4$, and $\{\widehat{\theta }_{0},\widehat{\theta }%
_{1},\widehat{\theta }_{2}\}=\{0^{\circ },138.98^{\circ },239.06^{\circ }\}$%
\end{tabular}
\\ 
\begin{tabular}{c||c|c}
\hline
Family & 
\begin{tabular}{c}
Family Size \\ 
($\Psi _{\text{max}}(N)$)%
\end{tabular}
& 
\begin{tabular}{c}
Factor Sets of $\widetilde{N}^{(0)}$, \\ 
$\widetilde{N}^{(1)}$, and $\widetilde{N}^{(2)}$%
\end{tabular}
\\ \hline\hline
$\widehat{\mathcal{G}}_{I}^{(\text{pma})},\widehat{\mathcal{G}}_{I}^{(\text{%
apma})}$ & 
\begin{tabular}{c}
$32,64$ \\ 
($567$)%
\end{tabular}
& 
\begin{tabular}{c}
$\{3,3,5,5\},\{2,2,$ \\ 
$7,7\},\{2,3,5,5\}$%
\end{tabular}
\\ \hline
$\widehat{\mathcal{G}}_{\text{max,}\widetilde{I}}^{(\text{dpma,}1)},\widehat{%
\mathcal{G}}_{\text{max,}\widetilde{I}}^{(\text{adpma,}1)}$ & 
\begin{tabular}{c}
$80,160$ \\ 
($569$)%
\end{tabular}
& 
\begin{tabular}{c}
$\{5,5,9\},\{4,7,$ \\ 
$7\},\{5,5,6\}$%
\end{tabular}
\\ \hline
$\widehat{\mathcal{G}}_{\text{max,}\widetilde{I}}^{(\text{dpma,}2)},\widehat{%
\mathcal{G}}_{\text{max,}\widetilde{I}}^{(\text{adpma,}2)}$ & 
\begin{tabular}{c}
$126,252$ \\ 
($569$)%
\end{tabular}
& 
\begin{tabular}{c}
$\{15,15\},\{14,$ \\ 
$14\},\{10,15\}$%
\end{tabular}
\\ \hline
\end{tabular}%
\end{tabular}

\begin{tabular}{c}
\begin{tabular}{c}
(c) $N=839$ with $(\widetilde{N}^{(0)},\widetilde{N}^{(1)},\widetilde{N}%
^{(2)})=(396,243,200)$, \\ 
$\widetilde{\Omega }(N)=5$, and $\{\widehat{\theta }_{0},\widehat{\theta }%
_{1},\widehat{\theta }_{2}\}=\{0^{\circ },156.04^{\circ },209.57^{\circ }\}$%
\end{tabular}
\\ 
\begin{tabular}{c||c|c}
\hline
Family & 
\begin{tabular}{c}
Family Size \\ 
($\Psi _{\text{max}}(N)$)%
\end{tabular}
& 
\begin{tabular}{c}
Factor Sets of $\widetilde{N}^{(0)},$ \\ 
$\widetilde{N}^{(1)},$ and $\widetilde{N}^{(2)}$%
\end{tabular}
\\ \hline\hline
$\widehat{\mathcal{G}}_{I}^{(\text{pma})},\widehat{\mathcal{G}}_{I}^{(\text{%
apma})}$ & 
\begin{tabular}{c}
$16,32$ \\ 
($835$)%
\end{tabular}
& 
\begin{tabular}{c}
$\{2,2,3,3,11\},$ \\ 
$\{3,3,3,3,3\},$ \\ 
$\{2,2,2,5,5\}$%
\end{tabular}
\\ \hline
$\widehat{\mathcal{G}}_{\text{max,}\widetilde{I}}^{(\text{dpma,}1)},\widehat{%
\mathcal{G}}_{\text{max,}\widetilde{I}}^{(\text{adpma,}1)}$ & 
\begin{tabular}{c}
$48,96$ \\ 
($835$)%
\end{tabular}
& 
\begin{tabular}{c}
$\{3,3,4,11\},\{3,3,$ \\ 
$3,9\},\{2,4,5,5\}$%
\end{tabular}
\\ \hline
$\widehat{\mathcal{G}}_{\text{max,}\widetilde{I}}^{(\text{dpma,}2)},\widehat{%
\mathcal{G}}_{\text{max,}\widetilde{I}}^{(\text{adpma,}2)}$ & 
\begin{tabular}{c}
$112,224$ \\ 
($837$)%
\end{tabular}
& 
\begin{tabular}{c}
$\{6,6,11\},$ \\ 
$\{3,9,9\},\{5,5,8\}$%
\end{tabular}
\\ \hline
$\widehat{\mathcal{G}}_{\text{max,}\widetilde{I}}^{(\text{dpma,}3)},\widehat{%
\mathcal{G}}_{\text{max,}\widetilde{I}}^{(\text{adpma,}3)}$ & 
\begin{tabular}{c}
$171,342$ \\ 
($837$)%
\end{tabular}
& 
\begin{tabular}{c}
$\{18,22\},$ \\ 
$\{9,27\},\{10,20\}$%
\end{tabular}
\\ \hline
\end{tabular}%
\end{tabular}

\begin{tabular}{c}
\begin{tabular}{c}
(d) $N=1151$ with $(\widetilde{N}^{(0)},\widetilde{N}^{(1)},\widetilde{N}%
^{(2)})=(468,440,243)$, \\ 
$\widetilde{\Omega }(N)=5$, and $\{\widehat{\theta }_{0},\widehat{\theta }%
_{1},\widehat{\theta }_{2}\}=\{0^{\circ },149.15^{\circ },248.20^{\circ }\}$%
\end{tabular}
\\ 
\begin{tabular}{c||c|c}
\hline
Family & 
\begin{tabular}{c}
Family Size \\ 
($\Psi _{\text{max}}(N)$)%
\end{tabular}
& 
\begin{tabular}{c}
Factor Sets of $\widetilde{N}^{(0)},$ \\ 
$\widetilde{N}^{(1)},$ and $\widetilde{N}^{(2)}$%
\end{tabular}
\\ \hline\hline
$\widehat{\mathcal{G}}_{I}^{(\text{pma})},\widehat{\mathcal{G}}_{I}^{(\text{%
apma})}$ & 
\begin{tabular}{c}
$32,64$ \\ 
($1147$)%
\end{tabular}
& 
\begin{tabular}{c}
$\{2,2,3,3,13\},$ \\ 
$\{2,2,2,5,11\},$ \\ 
$\{3,3,3,3,3\}$%
\end{tabular}
\\ \hline
$\widehat{\mathcal{G}}_{\text{max,}\widetilde{I}}^{(\text{dpma,}1)},\widehat{%
\mathcal{G}}_{\text{max,}\widetilde{I}}^{(\text{adpma,}1)}$ & 
\begin{tabular}{c}
$64,128$ \\ 
($1147$)%
\end{tabular}
& 
\begin{tabular}{c}
$\{3,3,4,13\},\{2,4,$ \\ 
$5,11\},\{3,3,3,9\}$%
\end{tabular}
\\ \hline
$\widehat{\mathcal{G}}_{\text{max,}\widetilde{I}}^{(\text{dpma,}2)},\widehat{%
\mathcal{G}}_{\text{max,}\widetilde{I}}^{(\text{adpma,}2)}$ & 
\begin{tabular}{c}
$128,256$ \\ 
($1149$)%
\end{tabular}
& 
\begin{tabular}{c}
$\{6,6,13\},\{5,8,$ \\ 
$11\},\{3,9,9\}$%
\end{tabular}
\\ \hline
$\widehat{\mathcal{G}}_{\text{max,}\widetilde{I}}^{(\text{dpma,}3)},\widehat{%
\mathcal{G}}_{\text{max,}\widetilde{I}}^{(\text{adpma,}3)}$ & 
\begin{tabular}{c}
$208,416$ \\ 
($1149$)%
\end{tabular}
& 
\begin{tabular}{c}
$\{18,26\},$ \\ 
$\{20,22\},\{9,27\}$%
\end{tabular}
\\ \hline
\end{tabular}%
\end{tabular}

\end{table}%

\emph{C)} \emph{Examples of Families }$\widehat{\mathcal{G}}_{I}^{(\text{apma%
})}$, $\widehat{\mathcal{G}}_{\text{max,}\widetilde{I}}^{(\text{dpma,}\kappa
)}$, \emph{and} $\widehat{\mathcal{G}}_{\text{max,}\widetilde{I}}^{(\text{%
adpma,}\kappa )}$\emph{: }The developed families $\widehat{\mathcal{G}}%
_{I}^{(\text{apma})}$, $\widehat{\mathcal{G}}_{\text{max,}\widetilde{I}}^{(%
\text{dpma,}\kappa )}$ and $\widehat{\mathcal{G}}_{\text{max,}\widetilde{I}%
}^{(\text{adpma,}\kappa )}$ are listed in Table II for example sequence
lengths adopted by the random access application in 5G-NR \cite{5G}, where
the achieved family size and the proper factor sets for all subsequence
lengths $\widetilde{N}^{(0)}$, $\widetilde{N}^{(1)}$,..., $\widetilde{N}%
^{(L-1)}$ are demonstrated for each family. As shown, family $\widehat{%
\mathcal{G}}_{\text{max,}\widetilde{I}}^{(\text{dpma,}\kappa )}$ offers the
larger family size as $\kappa $ increases and much larger family size than $%
\widehat{\mathcal{G}}_{I}^{(\text{pma})}$. Nevertheless, $\widehat{\mathcal{G%
}}_{\text{max,}\widetilde{I}}^{(\text{dpma,}\kappa )}$ may entail reduced SD
order $\widetilde{I}\geq \widetilde{\Omega }(N)-\kappa $ under \emph{%
Condition A} and $\widetilde{I}\geq \left\lfloor (\widetilde{\Omega }%
(N)-\kappa )/2\right\rfloor $ under \emph{Condition B}. For a fixed $\kappa
\in \mathcal{Z}_{\widetilde{\Omega }(N)-1}^{+}$, family $\widehat{\mathcal{G}%
}_{\text{max,}\widetilde{I}}^{(\text{adpma,}\kappa )}$ exhibits double the
size of family $\widehat{\mathcal{G}}_{\text{max,}\widetilde{I}}^{(\text{%
dpma,}\kappa )}$\ while sustaining the same SD order. Moreover, the family
sizes of $\widehat{\mathcal{G}}_{\text{max,}\widetilde{I}}^{(\text{dpma,}%
\kappa )}$ with large $\kappa $ values approach a good portion of the
largest achievable size $\Psi _{\text{max}}(N)$. The latter reveals the
advantage of augmented degenerate PMA sequence families.

\emph{Remark 1:} Orthogonal Zadoff-Chu sequences are desirable for the RA
application in 5G NR. Although $N$ orthogonal ZC sequences of length $N$ can
be easily generated by cyclically shifting the inverse DFT of a given ZC
sequence with an admissible root index $\varsigma $ (relatively prime to $N$%
), a large minimum CSD $\varpi _{\text{min}}$ is generally required to
identify received orthogonal ZC sequences transmitted from transmitters
located in various locations in the same cell. The larger the cell radius,
the larger the required $\varpi _{\text{min}}$. As a result, nonorthogonal
ZC sequences with different admissible root indices are commonly employed
for RA requiring a large number of short-length identification sequences
under a limited $\varpi _{\text{min}}$. In 5G NR, there are $64$ RA
identification sequences required in each cell, and many large values for $%
\varpi _{\text{min}}$ are specified in \cite[Tables 5-7 in Section 6.3.3.1]%
{5G} for the adopted ZC sequences of different lengths $N=139$, $571$, $839$%
, and $1151$. In these specifications, the maximum number of orthogonal ZC
sequences is limited to $\left\lfloor N/\varpi _{\text{min}}\right\rfloor <64
$ for many specified pairs $(N,\varpi _{\text{min}})$. For example with $%
(N,\varpi _{\text{min}})=(839,26)$, only $32$ orthogonal ZC sequences can be
generated by cyclically shifting the inverse DFT of a given ZC sequence with
an admissible root index $\varsigma _{1}$. In this case, additional
nonorthogonal ZC sequences are added in \cite[Section 6.3.3.1]{5G} by
cyclically shifting the inverse DFT of another ZC sequence with an
admissible root index $\varsigma _{2}$ so that all $64$ sequences are
collected. As shown in Table II(c), families $\widehat{\mathcal{G}}_{\text{%
max,}\widetilde{I}}^{(\text{adpma,}\kappa )}$ with $\kappa \in \mathcal{Z}%
_{3}^{+}$ and families $\widehat{\mathcal{G}}_{\text{max,}\widetilde{I}}^{(%
\text{dpma,}\kappa )}$ with $\kappa \in \{2,3\}$ can provide more than $64$
orthogonal order-$\widetilde{I}$ CA sequences and thus outperform the
adopted ZC sequences \cite[Section 6.3.3.1]{5G} in RA performance while
providing the higher spectral compactness.

\section{Random-Access Channel Identification}

This section demonstrates the performance characteristics of uplink RA
channel identification based on the reception of the OFDM preamble waveforms
carrying identification sequences from various CA sequence families,
including modified PMA, ZC, YL, and PN sequence families, over Rayleigh
multipath channels. Here, the interleaving factor $\gamma =1$ is considered.
Spectral compactness of various OFDM preamble waveforms are also shown to
justify the spectral compactness achieved by use of order-$I$ CA sequences.

Consider the scenario that a single user terminal transmits a sequence $%
\mathbf{q}_{k}\triangleq \lbrack q_{k}[n];n\in \mathcal{Z}_{N}]$ from the
family of $J$ CA sequences $\{\mathbf{q}_{i};i\in \mathcal{Z}_{J}\}$ for
identifying the availability of the $k$-th access channel \cite{LTE}-\cite%
{5G}. After applying down-conversion, CP removal, and DFT to the received
OFDM preamble signal, the basestation receiver observes the frequency-domain
vector $\mathbf{r}\triangleq \lbrack r[n];n\in \mathcal{Z}_{N}]$ modeled as 
\cite{RA 1}, \cite{RA 3}%
\begin{equation}
r[n]=N^{1/2}q_{k}[n]h[n]+z[n].  \label{received}
\end{equation}%
Here, $\mathbf{z}\triangleq \lbrack z[n];n\in \mathcal{Z}_{N}]$ contains
independent and identically distributed circularly symmetric complex
Gaussian (CSCG) noise samples with mean zero and variance $\mathcal{E}%
\{|z[n]|^{2}\}=1/\varphi $, where $\varphi $ is the received signal-to-noise
power ratio (SNR). $\mathbf{h}\triangleq \lbrack h[n];n\in \mathcal{Z}_{N}]$
is the channel frequency response (CFR) vector corresponding to the channel
impulse response (CIR) $\{\widetilde{h}[l],\tau _{l};l\in \mathcal{Z}%
_{L_{h}}\}$ with $L_{h}$ resolvable paths, given by%
\begin{equation}
h[n]=\sum\limits_{l\in \mathcal{Z}_{L_{h}}}\widetilde{h}[l]e^{-j2\pi
\bigtriangleup fn\tau _{l}}\text{ for all }n\in \mathcal{Z}_{N}  \label{cfr}
\end{equation}%
where $\bigtriangleup f=1/T_{\text{d}}$\ is subcarrier frequency spacing and 
$\tau _{l}$ denotes the $l$-th path delay value with $0\leq \tau _{0}<\tau
_{1}<...<\tau _{L_{h}-1}\leq T_{\text{g}}$. Moreover, all path gains $\{%
\widetilde{h}[l];l\in \mathcal{Z}_{L_{h}}\}$ are modeled to be independent
CSCGs having common mean zero and path powers $\mathcal{E}\{|\widetilde{h}%
[l]|^{2}\}=\sigma _{l}^{2}$ for $l$ $\in \mathcal{Z}_{L_{h}}$ with $%
\sum\nolimits_{l\in \mathcal{Z}_{L_{h}}}\sigma _{l}^{2}=1$, and also
independent of all noise samples $\{z[n];n\in \mathcal{Z}_{N}\}$. The RA
channel identification is based on the correlations $\{\mathbf{q}_{i}^{h}%
\mathbf{r};i\in \mathcal{Z}_{J}\}$, with%
\begin{eqnarray}
\mathbf{q}_{i}^{h}\mathbf{r} &=&N^{1/2}\sum\limits_{l\in \mathcal{Z}%
_{L_{h}}}\widetilde{h}[l]\sum\limits_{n\in \mathcal{Z}_{N}}q_{i}^{\ast
}[n]q_{k}[n]e^{-j2\pi \bigtriangleup fn\tau _{l}}  \notag \\
&&\mathbf{+}\sum\limits_{n\in \mathcal{Z}_{N}}q_{i}^{\ast }[n]z[n].
\end{eqnarray}%
To identify $\mathbf{q}_{k}$, the squared correlation magnitudes $Y(\mathbf{q%
}_{i})=|\mathbf{q}_{i}^{h}\mathbf{r}|^{2}$ are measured and compared with a
positive threshold $\beta $ for all $i\in \mathcal{Z}_{J}$. When $Y(\mathbf{q%
}_{i})$ is greater than $\beta $, the $i$-th access channel is considered as
a requested one \cite{RA 1}-\cite{RA 2}, \cite{RA 4}-\cite{RA 5}.

For $i\in \mathcal{Z}_{J}$, $\mathbf{q}_{i}^{h}\mathbf{r}$ is a CSCG having
zero mean and variance $\mathcal{E}\{|\mathbf{q}_{i}^{h}\mathbf{r}|^{2}\}=%
\frac{1}{\varphi }+\sigma _{\text{fie}}^{2}(i,k)$ if $i\neq k$ and $\mathcal{%
E}\{|\mathbf{q}_{i}^{h}\mathbf{r}|^{2}\}=\frac{1}{\varphi }+\sigma _{\text{c}%
}^{2}$ otherwise, where $\sigma _{\text{fie}}^{2}(i,k)\triangleq
N\sum\nolimits_{l\in \mathcal{Z}_{L_{h}}}$ $\sigma
_{l}^{2}|\sum\nolimits_{n\in \mathcal{Z}_{N}}q_{i}^{\ast
}[n]q_{k}[n]e^{-j2\pi \bigtriangleup fn\tau _{l}}|^{2}$ is the variance of
the FIE term occurring when $\mathbf{q}_{i}$ does not match the
identification sequence $\mathbf{q}_{k}$ and $\sigma _{\text{c}}^{2}=\frac{1%
}{N}\sum\nolimits_{l\in \mathcal{Z}_{L_{h}}}$ $\sigma
_{l}^{2}|\sum\nolimits_{n\in \mathcal{Z}_{N}}e^{-j2\pi \bigtriangleup
fn\tau _{l}}|^{2}$ is the signaling variance when $\mathbf{q}_{i}$ matches $%
\mathbf{q}_{k}$ correctly. Given the statistic of $\mathbf{q}_{i}^{h}\mathbf{%
r}$, $Y(\mathbf{q}_{i})$ is a central chi-square random variable with two
degrees of freedom \cite{Proakis}.

\begin{table}\centering%
\caption{A Typical RA System Parameter Profile in Uplink 5G NR.}%
\begin{tabular}{c||c}
\hline
Sequence Length $N$ & $839$ \\ \hline
Subcarrier Spacing $\bigtriangleup f=1/T_{\text{d}}$ & $1.25$ kHz \\ \hline
Interleaving Factor $\gamma $ & $1$ \\ \hline
Guard Ratio $\alpha $ & $33/256$ \\ \hline
Total Number of Sequences $J$ & $64$ \\ \hline
\end{tabular}%
\end{table}%

Three measures $P_{\text{fa}}$, $P_{\text{fid,}k}$, and $P_{\text{c}}$ are
defined herein to quantify the performance of the threshold-based
identification scheme. The false alarm probability $P_{\text{fa}}$ denotes
the probability of misidentifying $\mathbf{q}_{i}$ when there is no request
(i.e., $r[n]=z[n]$ for all $n\in \mathcal{Z}_{N}$), defined by $P_{\text{fa}%
}\triangleq \Pr \{Y(\mathbf{q}_{i})>\beta |$no request$\}$ for some $i\in 
\mathcal{Z}_{J}$ and given by $P_{\text{fa}}=e^{-\beta \varphi }$, which is
invariant with $\mathbf{q}_{i}$. The average false identification
probability $P_{\text{fid,}k}$ is the average probability of identifying the
request of an access channel other than the $k$-th channel that was actually
requested \cite[Subsection IV.\emph{D}]{RA 4}, and given by%
\begin{eqnarray}
P_{\text{fid,}k} &\triangleq &\frac{1}{J-1}\sum\limits_{i\in \mathcal{Z}%
_{J},i\neq k}\Pr \{Y(\mathbf{q}_{i})>\beta |\mathbf{q}_{k}\text{ was
requested}\}  \notag \\
&=&\frac{1}{J-1}\sum\limits_{i\in \mathcal{Z}_{J},i\neq k}e^{-\beta \varphi
/(1+\varphi \sigma _{\text{fie}}^{2}(i,k))}.  \label{fid}
\end{eqnarray}%
From the union bound argument, $(J-1)P_{\text{fid,}k}$ is also an upper
bound to the probability of identifying the request of \emph{any} access
channel other than the $k$-th channel that was actually requested \cite[%
Subsection III.\emph{D}]{RA 5}. The correct identification probability $P_{%
\text{c}}$ is the average probability of identifying the request of the $k$%
-th access channel correctly, defined by $P_{\text{c}}\triangleq \Pr \{Y(%
\mathbf{q}_{k})>\beta |\mathbf{q}_{k}$ was requested$\}$ and given by $P_{%
\text{c}}=e^{-\beta \varphi /(1+\varphi \sigma _{\text{c}}^{2})}$, which is
irrelevant with $\mathbf{q}_{k}$. The identification scheme performs well
when $P_{\text{c}}$ is made as large as possible while $P_{\text{fa}}$ and
all $P_{\text{fid,}k}$ are restricted to be small. This can be achieved by
properly setting the threshold $\beta $ since $P_{\text{fa}}$, $P_{\text{fid,%
}k}$, and $P_{\text{c}}$ increase as $\beta $ is decreased for a given SNR $%
\varphi $. When the channel is flat fading (i.e., $h[n]=\widetilde{h}[0]$
for all $n\in \mathcal{Z}_{N}$, or equivalently $L_{h}=1$, $\tau _{0}=0$,
and $\sigma _{0}^{2}=1$), $\mathbf{q}_{i}^{h}\mathbf{r}$ for $i\neq k$
simplifies to a CSCG with mean zero and variance $\mathcal{E}\{|\mathbf{q}%
_{i}^{h}\mathbf{r}|^{2}\}=\frac{1}{\varphi }+\widetilde{\sigma }_{\text{fie}%
}^{2}(i,k)$, where $\widetilde{\sigma }_{\text{fie}}^{2}(i,k)=N|\sum%
\nolimits_{n\in \mathcal{Z}_{N}}q_{i}^{\ast }[n]q_{k}[n]|^{2}$. In this
case, $P_{\text{fid,}k}$ in (\ref{fid}) achieves the minimum $P_{\text{%
fid,min}}=e^{-\beta \varphi }$ when all sequences in the family $\{\mathbf{q}%
_{i};i\in \mathcal{Z}_{J}\}$ are mutually orthogonal. Moreover, $P_{\text{c}%
} $ achieves the maximum $P_{\text{c,max}}=e^{-\beta \varphi /(1+\varphi N)}$%
. When the coherence bandwidth $B_{\text{c}}\approx 1/(5\sigma _{\text{rms}%
}) $ \cite[Chapter 4, eq. 39]{Coherence bandwidth} is much larger than the
signaling bandwidth $N\gamma /T_{d}$ (i.e., $\tau _{0}=0$ and $\sigma
_{l}^{2}\ll \sigma _{0}^{2}$ for all $l\neq 0$), $\sigma _{\text{fie}%
}^{2}(i,k)$ approaches to $\widetilde{\sigma }_{\text{fie}}^{2}(i,k)$ for $%
i\neq k$ and $P_{\text{fid,}k}$ is expected to get close to $P_{\text{fid,min%
}}$ if all sequences in $\{\mathbf{q}_{l};l\in \mathcal{Z}_{J}\}$ are
orthogonal, where $\sigma _{\text{rms}}$ is the root mean square delay
spread in CIR. As thus implied, smaller $P_{\text{fid,}k}$ values can be
achieved when there are more orthogonal sequences in $\{\mathbf{q}_{i};i\in 
\mathcal{Z}_{J}\}$ available for RA channel identification over multipath
channels with large coherence bandwidth, or equivalently short-delay channel
profiles.

\begin{figure}[h]%
\centering
\includegraphics[width=0.48\textwidth]{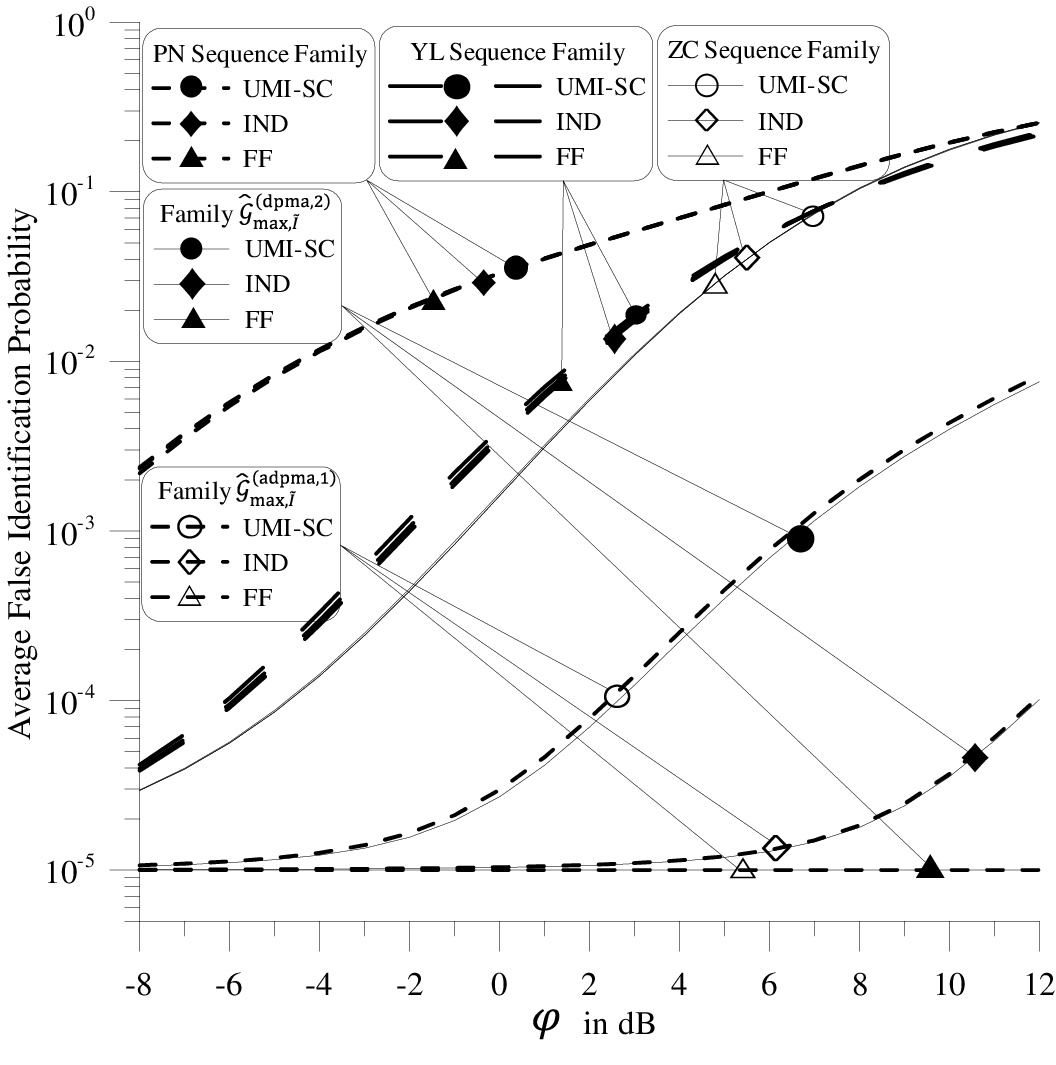}
\caption{The characteristics of the
average false identification probability versus SNR among the RA channel
identification systems using various sequence families under TDL-B UMI-SC
short-delay channel profile, TDL-B IND short-delay channel profile, and FF
channel profile.}\label{fig1}
\end{figure}

A total of $64$ ZC sequences are required for RA channel identification in
uplink 5G-NR \cite[Section 6.3.3.1]{5G}. To avoid sequence identification
ambiguity, a minimum CSD $\varpi _{\text{min}}$ is required to extract
cyclically-shiftable ZC sequences through cyclically shifting the inverse
DFT of a single-root ZC sequence. As mentioned in \emph{Remark 1}, this
causes the shortage of adoptable cyclically-shiftable ZC sequences for most
specified $(N,\varpi _{\text{min}})$ pairs. In \cite{YL sequence},
orthogonal YL sequences are constructed from phase-rotating the ZC sequences
generated from cyclically shifting the inverse DFT of a single-root ZC
sequence appropriately. When the minimum CSD requirement is imposed, not
every cyclically-shiftable ZC sequence can be used to generate orthogonal YL
sequences. The latter limits the number of adoptable orthogonal YL sequences
as well in order to avoid sequence identification ambiguity. For example, we
consider a particular RA system parameter profile in Table III \cite{5G}
which adopts the sequence length $N=839$ and the minimum CSD limit $26$. In
this case, at most $\left\lfloor 839/26\right\rfloor =32$
cyclically-shiftable ZC and YL sequences can be respectively adopted and
thus nonorthogonal sequences have to be augmented in \cite[Section 6.3.3.1]%
{5G} since $64$ RA channels are to be identified. The characteristics of the
average false identification probability $\frac{1}{J}\sum\nolimits_{k\in 
\mathcal{Z}_{J}}P_{\text{fid,}k}$ versus SNR $\varphi $ are demonstrated in
Fig. 1 by simulating the threshold-based RA channel identification using
such ZC and YL sequence families under three different channel profiles,
namely TDL-B urban micro street-canyon (UMI-SC) short-delay profile
(exhibiting $\sigma _{\text{rms}}=65$ ns and $B_{\text{c}}\approx 2.93\times
N\gamma /T_{\text{d}}$) and TDL-B indoor (IND) short-delay profile
(exhibiting $\sigma _{\text{rms}}=20$ ns and $B_{\text{c}}\approx 9.54\times
N\gamma /T_{\text{d}}$) in \cite[Section 7.7.2]{5G Channel} as well as the
benchmarking flat fading (FF) channel profile (exhibiting an infinitely
large $B_{\text{c}}$). Notably, the UMI-SC short-delay profile exhibits a
longer delay spread than the IND short-delay profile, and thus results in a
smaller coherence bandwidth. Also compared in Fig. 1 are RA channel
identification systems using orthogonal sequence families $\widehat{\mathcal{%
G}}_{\text{max,}\widetilde{I}}^{(\text{adpma,}1)}$ and $\widehat{\mathcal{G}}%
_{\text{max,}\widetilde{I}}^{(\text{dpma,}2)}$, and a nonorthogonal PN
sequence family. All $64$ PN sequences are constructed from the generator
polynomial $X^{15}+X^{14}+1$ with minimum CSD $26$ \cite[Section 9.7.1]{Wifi}%
. As described in Table II(c), families $\widehat{\mathcal{G}}_{\text{max,}%
\widetilde{I}}^{(\text{adpma,}1)}$ and $\widehat{\mathcal{G}}_{\text{max,}%
\widetilde{I}}^{(\text{dpma,}2)}$ can provide $96$ and $112$ orthogonal
sequences, respectively, and $64$ sequences are randomly chosen from them in
the simulation. To achieve an extremely small $P_{\text{fa}}=10^{-5}$, the
threshold value is set to $\beta =\frac{5}{\varphi }\ln 10$ for a given SNR $%
\varphi $ and in this case the correct identification probability is
equivalent to $P_{\text{c}}=10^{-5/(1+\varphi \sigma _{\text{c}}^{2})}$. For
the SNR range demonstrated in Fig. 1, $1-P_{\text{c}}$ falls in the ranges $%
[8.85\times 10^{-4},8.42\times 10^{-2}]$, $[8.67\times 10^{-4},8.25\times
10^{-2}]$, and $[8.65\times 10^{-4},8.23\times 10^{-2}]$ for UMI-SC, IND,
and FF channel profiles, respectively. Due to the adoption of nonorthogonal
sequences, RA channel identification suffers from large FIE (i.e., larger $%
\sigma _{\text{fie}}^{2}(i,k)$) and thus entails serious false
identification for the systems using ZC, YL, and PN sequence families. On
the contrary, false identification is less severe for the systems using
orthogonal sequence families $\widehat{\mathcal{G}}_{\text{max,}\widetilde{I}%
}^{(\text{dpma,}2)}$ and $\widehat{\mathcal{G}}_{\text{max,}\widetilde{I}}^{(%
\text{adpma,}1)}$, particularly in the multipath channels exhibiting larger
coherence bandwidths.

Fig. 2 compares the spectral compactness characteristics of all the OFDM
preamble waveforms adopted in Fig. 1. To compare the spectral compactness
among various waveforms, the average out-of-band power fraction is defined as%
\begin{equation*}
\eta \triangleq 10\log _{10}(\frac{1}{J}\sum\limits_{i\in \mathcal{Z}%
_{J}}(\int\nolimits_{|f|>B/2}S_{B}^{(i)}(f)df/\int\nolimits_{-\infty
}^{\infty }S_{B}^{(i)}(f)df))
\end{equation*}%
where $S_{B}^{(i)}(f)$ is the baseband power spectrum of the waveform
carrying $\mathbf{q}_{i}$ \cite{CA1}. The results on $\eta $ are presented
with respect to the normalized bandwidth $BT_{\text{d}}/(\gamma N)$. For a
predetermined $\eta $ (say $-50$ dB), the smaller the required bandwidth,
the higher the spectral compactness. As shown, preamble waveforms carrying
order-$\widetilde{I}$ CA sequence families $\widehat{\mathcal{G}}_{\text{max,%
}\widetilde{I}}^{(\text{adpma,}1)}$ (yielding SD order $\widetilde{I}\geq 2$%
) and $\widehat{\mathcal{G}}_{\text{max,}\widetilde{I}}^{(\text{dpma,}2)}$
(yielding SD order $\widetilde{I}\geq 1$) can provide much higher spectral
compactness than preamble waveforms carrying ZC, YL, PN sequence families.%
\begin{figure}[h]%
\centering
\includegraphics[width=0.48\textwidth]{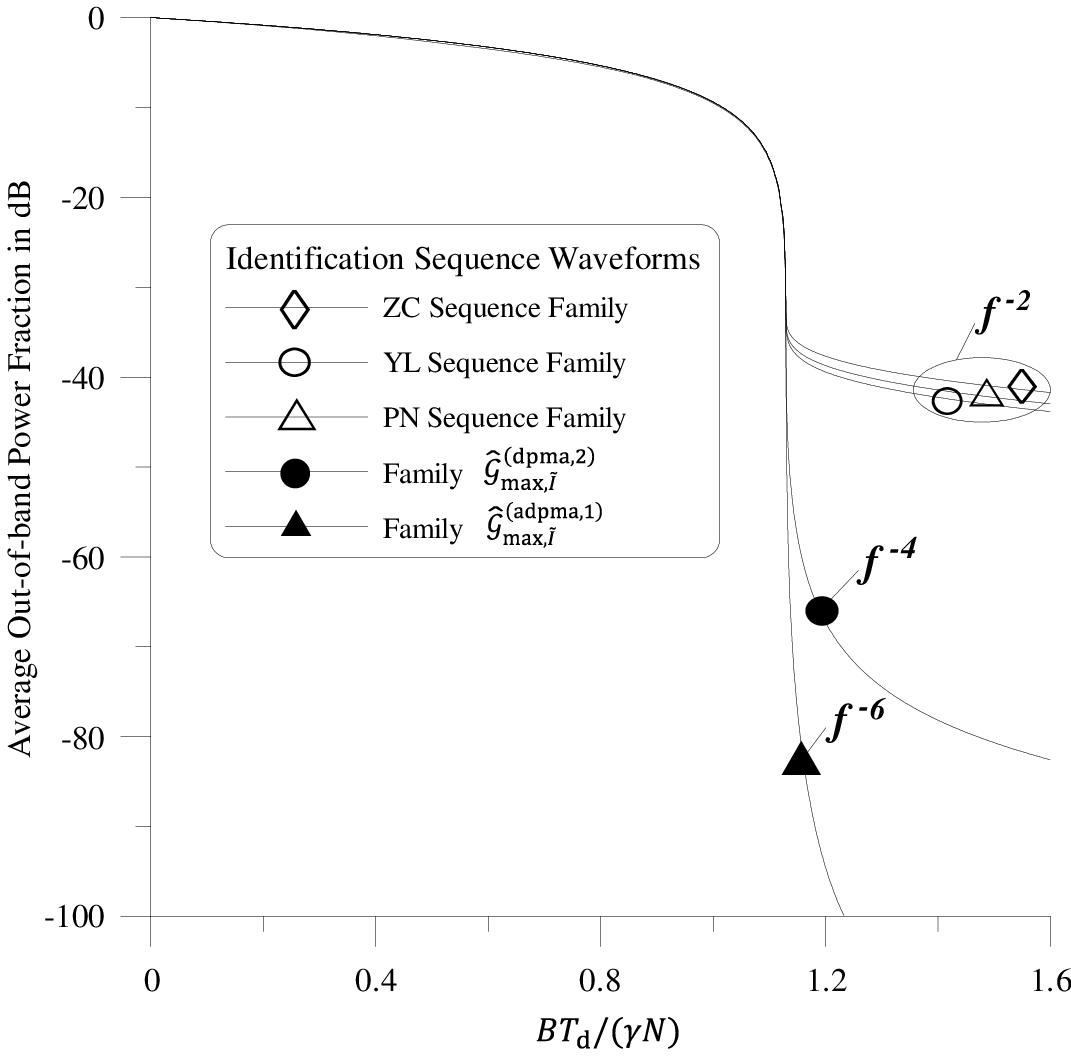}
\caption{Average out-of-band power
fraction characteristics for OFDM preamble waveforms carrying various CA
sequence families.}\label{fig2}
\end{figure}

\section{Conclusion}

Several modified PMA sequence families are constructed in the paper to
provide more orthogonal order-$I$ CA sequences for SPI applications, while
facilitating the composition of spectrally compact OFDM preamble/pilot
waveforms. The higher the sidelobe-decaying order, the higher spectral
compactness the preamble/pilot waveform exhibits. By use of the developed
orthogonal order-$I$ CA sequences, the SPI system requiring a large number
of identification/sounding sequences can achieve the better performance in
multipath channels exhibiting short-delay channel profiles, while exhibiting
high spectral compactness. Specifically, degenerate PMA sequence families $%
\mathcal{G}_{\text{max,}\widetilde{I}}^{(\text{dpma,}\kappa )}$, $\widetilde{%
\mathcal{G}}_{\text{max,}\widetilde{I}}^{(\text{dpma,}\kappa )}$, and $%
\widehat{\mathcal{G}}_{\text{max,}\widetilde{I}}^{(\text{dpma,}\kappa )}$
are constructed by properly factorizing the sequence length in the
construction of PMA sequences with or without reducing the sidelobe-decaying
order. Augmented PMA sequence families $\widehat{\mathcal{G}}_{I}^{(\text{%
apma})}$ and $\widehat{\mathcal{G}}_{\text{max,}\widetilde{I}}^{(\text{adpma,%
}\kappa )}$ are further constructed to double the family size by augmenting
the phase-rotated replicas of all PMA sequences in families $\widehat{%
\mathcal{G}}_{I}^{(\text{pma})}$ and $\widehat{\mathcal{G}}_{\text{max,}%
\widetilde{I}}^{(\text{dpma,}\kappa )}$, respectively, without trading off
the sidelobe-decaying order. When compared with conventional Zadoff-Chu,
Yu-Lee, and pseudorandom-noise CA sequence families, these modified PMA
sequence families are shown to provide noticeable performance improvement in
random-access channel identification over indoor and urban multipath
environments exhibiting short-delay channel profiles. Meanwhile, the
preamble/pilot waveforms carrying order-$I$ CA sequences in these modified
PMA sequence families are attributed with much higher spectral compactness
than those carrying conventional CA sequences.


\section*{APPENDIX}

\emph{A) Proof of Lemma 1:} Consider two $M$-tuples $\mathbf{1}_{m}$ and $%
\mathbf{x}=[x_{m};m\in \mathcal{Z}_{M}]$ where all entries $x_{m}$ are
integers greater than one and $\mathbf{1}_{m}$ contains one at the $m$-th
entry and $M-1$ zeros elsewhere. With (\ref{11}), $f(\mathbf{x}^{t}+k_{m}%
\mathbf{1}_{m}^{t})-f(\mathbf{x}^{t})$ is given by%
\begin{eqnarray*}
&&f(\mathbf{x}^{t}+k_{m}\mathbf{1}_{m}^{t})-f(\mathbf{x}^{t}) \\
&=&\frac{(x_{m}+k_{m})\prod\nolimits_{i\neq m}x_{i}-1}{(x_{m}+k_{m}-1)%
\prod\nolimits_{i\neq m}(x_{i}-1)}-\frac{\prod\nolimits_{i}x_{i}-1}{%
\prod\nolimits_{i}(x_{i}-1)} \\
&=&\frac{1}{\prod\nolimits_{i\neq m}(x_{i}-1)}[\frac{(x_{m}+k_{m})\prod%
\nolimits_{i\neq m}x_{i}-1}{x_{m}+k_{m}-1}-\frac{\prod\nolimits_{i}x_{i}-1}{%
x_{m}-1}] \\
&=&\frac{k_{m}(1-\prod\nolimits_{i\neq m}x_{i})}{(x_{m}+k_{m}-1)\prod%
\nolimits_{i}(x_{i}-1)}
\end{eqnarray*}%
for $m\in \mathcal{Z}_{M}$ and it is negative when $k_{m}$ is a positive
integer. Thus, $f(\mathbf{x}^{t}+k_{m}\mathbf{1}_{m}^{t})<f(\mathbf{x}^{t})$
if the integer $k_{m}$ is positive and obviously $f(\mathbf{x}^{t}+k_{m}%
\mathbf{1}_{m}^{t})=f(\mathbf{x}^{t})$ if $k_{m}=0$.

Next, define another $M$-tuple $\mathbf{k}=\mathbf{b}-\mathbf{a}$ and
express $\mathbf{b}$ in terms of $\mathbf{a}$ and $\mathbf{k}$ as%
\begin{equation*}
\mathbf{b}=\mathbf{a}+\mathbf{k}=\mathbf{a}+\sum\nolimits_{m=0}^{M-1}k_{m}%
\mathbf{1}_{m}
\end{equation*}%
where all integer-valued entries $k_{m}$ in $\mathbf{k}=[k_{m};m\in \mathcal{%
Z}_{M}]$ are nonnegative and all integer-valued entries $a_{m}$ and $b_{m}$
in $\mathbf{a}=[a_{m};m\in \mathcal{Z}_{M}]$ and $\mathbf{b}=[b_{m};m\in 
\mathcal{Z}_{M}]$ are greater than one. With $f(\mathbf{x}^{t}+k_{m}\mathbf{1%
}_{m}^{t})<f(\mathbf{x}^{t})$ for a positive $k_{m}$, we have%
$
f(\mathbf{b}^{t})\leq f(\mathbf{a}^{t}+\sum\nolimits_{m=0}^{M-2}k_{m}%
\mathbf{1}_{m}^{t})\leq ...\leq f(\mathbf{a}^{t}+k_{0}\mathbf{1}%
_{0}^{t})\leq f(\mathbf{a}^{t}).$
Thus, $f(\mathbf{a}^{t})\geq f(\mathbf{b}^{t})$ if $1<a_{n}\leq b_{n}$ for
all $n\in \mathcal{Z}_{M}$, and $f(\mathbf{a}^{t})>f(\mathbf{b}^{t})$ if $%
1<a_{n}<b_{n}$ for some $n\in \mathcal{Z}_{M}$ and $1<a_{m}\leq b_{m}$ for
all $m\in \mathcal{Z}_{M}-\{n\}$. This completes the proof.

\emph{B) Proof of Lemma 2:} With (\ref{11}), $f([P_{a},P_{d}])\times
f([P_{b},P_{c}])-f([P_{a},P_{c}])\times f([P_{b},P_{d}])$ is given by%
\begin{eqnarray}
&&\frac{(P_{a}P_{d}-1)(P_{b}P_{c}-1)-(P_{a}P_{c}-1)(P_{b}P_{d}-1)}{%
(P_{a}-1)(P_{b}-1)(P_{c}-1)(P_{d}-1)}  \notag \\
&=&\frac{P_{a}P_{c}+P_{b}P_{d}-P_{a}P_{d}-P_{b}P_{c}}{%
(P_{a}-1)(P_{b}-1)(P_{c}-1)(P_{d}-1)}  \notag \\
&=&\frac{(P_{b}-P_{a})(P_{d}-P_{c})}{(P_{a}-1)(P_{b}-1)(P_{c}-1)(P_{d}-1)}.
\label{13}
\end{eqnarray}%
Similarly, $f([P_{a},P_{c}])\times f([P_{b},P_{d}])-f([P_{a},P_{b}])\times
f([P_{c},P_{d}])$ is given by%
\begin{equation}
\frac{(P_{d}-P_{a})(P_{c}-P_{b})}{(P_{a}-1)(P_{b}-1)(P_{c}-1)(P_{d}-1)}.
\label{14}
\end{equation}%
When $1<P_{a}\leq P_{b}\leq P_{c}\leq P_{d}$, (\ref{13}) and (\ref{14}) are
both nonnegative. This completes the proof.
\begin{figure}[h]%
\centering
\includegraphics[width=0.48\textwidth]{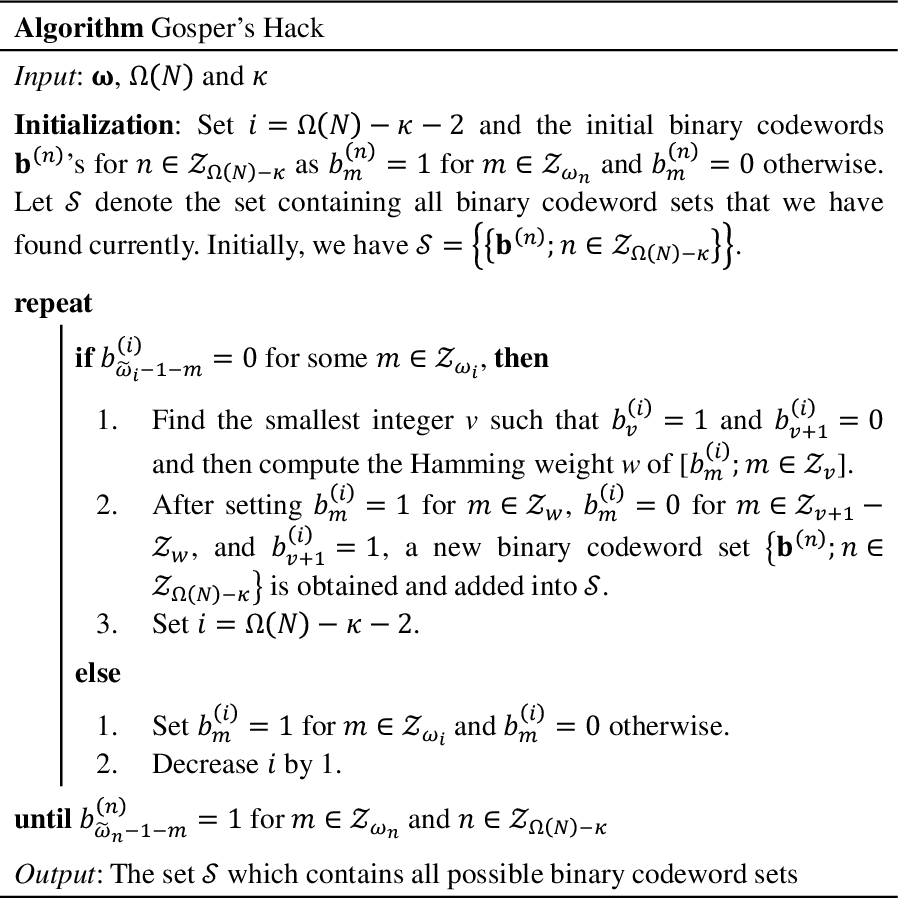}
\caption{Gosper's Hack algorithm.}\label{fig3}
\end{figure}

\emph{C) Proof of Lemma 3:} With (\ref{11}), $%
f([P_{a},P_{b},P_{c}])-f([P_{a},P_{d}])\times f([P_{b},P_{c}])$ is given by%
\begin{eqnarray}
&&\frac{P_{a}P_{b}P_{c}-1}{(P_{a}-1)(P_{b}-1)(P_{c}-1)}  \notag \\
&&-\frac{(P_{a}P_{d}-1)(P_{b}P_{c}-1)}{(P_{a}-1)(P_{b}-1)(P_{c}-1)(P_{d}-1)}
\notag \\
&=&\frac{P_{a}P_{d}+P_{b}P_{c}-P_{a}P_{b}P_{c}-P_{d}}{%
(P_{a}-1)(P_{b}-1)(P_{c}-1)(P_{d}-1)}  \notag \\
&=&\frac{(P_{a}-1)(P_{d}-P_{b}P_{c})}{(P_{a}-1)(P_{b}-1)(P_{c}-1)(P_{d}-1)}
\label{15}
\end{eqnarray}%
which is nonnegative when $P_{b}P_{c}\leq P_{d}$ and $1<P_{a}\leq P_{b}\leq
P_{c}\leq P_{d}$. This completes the proof.

\emph{D) Gosper's Hack Algorithm:} Gosper's Hack algorithm in \cite{Gosper's
Hack}-\cite{Gospe's Hack Alg} can assist in finding all possible factor sets 
$\{A_{m};m\in \mathcal{Z}_{\Omega (N)-\kappa }\}$ which satisfy $%
\prod\nolimits_{m=0}^{\Omega (N)-\kappa
-1}A_{m}=\prod\nolimits_{m=0}^{\Omega (N)-1}P_{m}$ and are all
characterized by an admissible pattern $\bm{\omega }=[\omega _{m};m\in 
\mathcal{Z}_{\Omega (N)-\kappa }]$ with $\omega _{m}=\Omega (A_{m})$. To
find all possible factor sets $\{A_{m};m\in \mathcal{Z}_{\Omega (N)-\kappa
}\}$, we aim to (i) first find all possible partitions of $\{P_{m};m\in 
\mathcal{Z}_{\Omega (N)}\}$ into $\Omega (N)-\kappa $ prime factor subsets $%
\{P_{m}^{(n)};m\in \mathcal{Z}_{\omega _{n}}\}$ for $n\in \mathcal{Z}%
_{\Omega (N)-\kappa }$, where $P_{0}^{(n)}\leq P_{1}^{(n)}\leq ...\leq
P_{\omega _{n}-1}^{(n)}$, with the aid of Gosper's Hack algorithm and (ii)
then compose all possible factor sets by computing $A_{n}=\prod%
\nolimits_{m=0}^{\omega _{n}-1}P_{m}^{(n)}$ accordingly. To describe step
(i), we define $\widetilde{\bm{\omega }}=[\widetilde{\omega }_{n};n\in 
\mathcal{Z}_{\Omega (N)-\kappa }]$ with $\widetilde{\omega }_{n}\triangleq
\sum\nolimits_{m=n}^{\Omega (N)-\kappa -1}\omega _{m}$ and $\mathbf{b}%
^{(n)}\triangleq \lbrack b_{m}^{(n)};m\in \mathcal{Z}_{\widetilde{\omega }%
_{n}}]$ as a binary codeword with length $\widetilde{\omega }_{n}$ and
Hamming weight $\omega _{n}$.\footnote{%
Notably, $\widetilde{\omega }_{0}=\Omega (N)$ and all Hamming weights $%
\omega _{m}$ sum to $\Omega (N)$.} For a given $\bm{\omega }$, there are
a total of $\prod\nolimits_{n\in \mathcal{Z}_{\Omega (N)-\kappa }}\tbinom{%
\widetilde{\omega }_{n}}{\omega _{n}}$ possible binary codeword sets for $\{%
\mathbf{b}^{(n)};n\in \mathcal{Z}_{\Omega (N)-\kappa }\}$ and they can be
exclusively obtained by Gosper's Hack algorithm in Fig. 3 \cite[Algorithm 3.1%
]{Gospe's Hack Alg}. To obtain a partition of $\{P_{m};m\in \mathcal{Z}%
_{\Omega (N)}\}$ for each given $\{\mathbf{b}^{(n)};n\in \mathcal{Z}_{\Omega
(N)-\kappa }\}$, a binary codeword set $\{\widetilde{\mathbf{b}}^{(n)};n\in 
\mathcal{Z}_{\Omega (N)-\kappa }\}$ is converted from $\{\mathbf{b}%
^{(n)};n\in \mathcal{Z}_{\Omega (N)-\kappa }\}$ by the proposed codeword
conversion algorithm in Fig. 4, in a way that each codeword $\widetilde{%
\mathbf{b}}^{(n)}\triangleq \lbrack \widetilde{b}_{m}^{(n)};m\in \mathcal{Z}%
_{\Omega (N)}]$ contains $\Omega (N)$ entries and the same Hamming weight as 
$\mathbf{b}^{(n)}$. Notably, there are a total of $\Omega (N)$ ones in $\{%
\widetilde{\mathbf{b}}^{(n)};n\in \mathcal{Z}_{\Omega (N)-\kappa }\}$. From $%
\{\widetilde{\mathbf{b}}^{(n)};n\in \mathcal{Z}_{\Omega (N)-\kappa }\}$, a
partition of $\{P_{m};m\in \mathcal{Z}_{\Omega (N)}\}$ into $\Omega
(N)-\kappa $ prime factor subsets $\{P_{\widetilde{m}}^{(n)};\widetilde{m}%
\in \mathcal{Z}_{\omega _{n}}\}$ can be thus specified by%
\begin{equation}
P_{\varepsilon _{m}^{(n)}}^{(n)}=P_{m}\text{ if }\widetilde{b}_{m}^{(n)}=1
\end{equation}%
for $n\in \mathcal{Z}_{\Omega (N)-\kappa }$ and $m\in \Omega (N)$, where $%
\varepsilon _{m}^{(n)}=\sum\nolimits_{m^{\prime }=0}^{m}\widetilde{b}%
_{m^{\prime }}^{(n)}-1$. Accordingly, all possible partitions of $%
\{P_{m};m\in \mathcal{Z}_{\Omega (N)}\}$ and thereby all possible factor
sets for $\{A_{m};m\in \mathcal{Z}_{\Omega (N)-\kappa }\}$ can be found in
steps (i) and (ii) from $\prod\nolimits_{n\in \mathcal{Z}_{\Omega
(N)-\kappa }}\tbinom{\widetilde{\omega }_{n}}{\omega _{n}}$ possible
codeword sets for $\{\mathbf{b}^{(n)};n\in \mathcal{Z}_{\Omega (N)-\kappa
}\} $.
\begin{figure}[h]%
\centering
\includegraphics[width=0.48\textwidth]{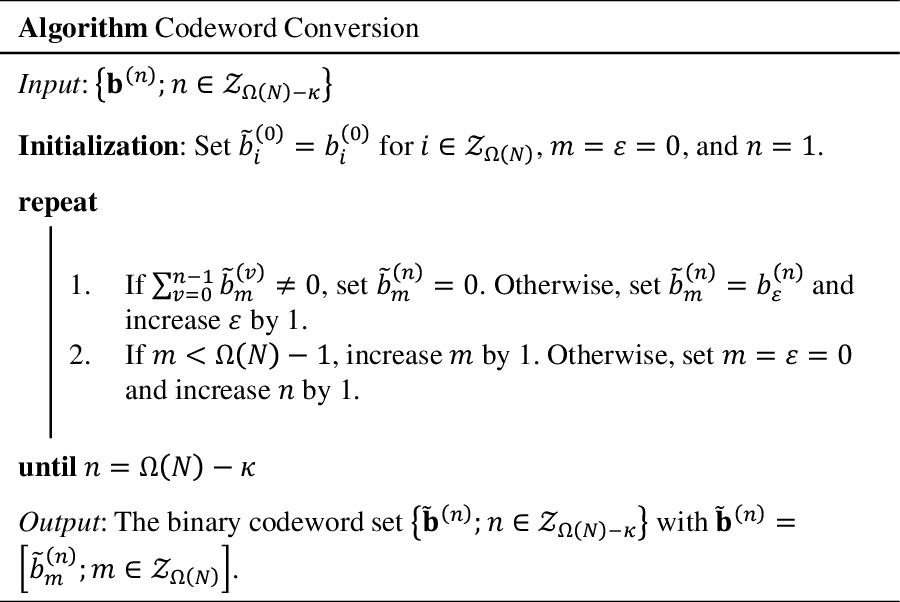}
\caption{Codeword conversion algorithm.}\label{fig4}
\end{figure}

Consider the example with $\Omega (N)=6$, $\kappa =3$, and a given pattern $%
\bm{\omega }=[3,2,1]^{t}$. Such $\bm{\omega }$ determines $%
\widetilde{\bm{\omega }}=$ $[6,3,1]^{t}$ uniquely and thus fixes the
lengths $6,3,1$ and Hamming weights $3,2,1$ of the binary codeword set $\{%
\mathbf{b}^{(0)},\mathbf{b}^{(1)},\mathbf{b}^{(2)}\}$ accordingly. From
Gosper's Hack algorithm, there are $\tbinom{6}{3}\tbinom{3}{2}\tbinom{1}{1}%
=60$ possible codeword sets meeting such length and weight distributions.
For example, $\mathbf{b}^{(0)}=[0,1,0,$ $1,1,0]^{t}$, $\mathbf{b}%
^{(1)}=[0,1,1]^{t}$ and $\mathbf{b}^{(2)}=[1]$ form one possible codeword
set. From the codeword conversion algorithm, the corresponding codeword set $%
\{\widetilde{\mathbf{b}}^{(n)};n\in \mathcal{Z}_{\Omega (N)-\kappa }\}$ is
obtained as $\widetilde{\mathbf{b}}^{(0)}=[0,1,0,1,1,0]^{t}$, $\widetilde{%
\mathbf{b}}^{(1)}=[0,0,1,0,0,1]^{t}$ and $\widetilde{\mathbf{b}}%
^{(2)}=[1,0,0,0,0,0]^{t}$. In turns, such $\{\widetilde{\mathbf{b}}%
^{(n)};n\in \mathcal{Z}_{\Omega (N)-\kappa }\}$ determines a partition of $%
\{P_{m};m\in \mathcal{Z}_{\Omega (N)}\}$ into $\{P_{m}^{(0)};m\in \mathcal{Z}%
_{\omega _{0}}\}=\{P_{1},P_{3},P_{4}\}$, $\{P_{m}^{(1)};m\in \mathcal{Z}%
_{\omega _{1}}\}=\{P_{2},P_{5}\}$, and $\{P_{m}^{(2)};m\in \mathcal{Z}%
_{\omega _{2}}\}=\{P_{0}\}$. The corresponding $\{A_{m};m\in \mathcal{Z}%
_{\Omega (N)-\kappa }\}$ becomes $\{P_{1}P_{3}P_{4},P_{2}P_{5},P_{0}\}$. All 
$60$ possible partitions can be thus obtained from $60$ codeword sets $\{%
\mathbf{b}^{(0)},$ $\mathbf{b}^{(1)},\mathbf{b}^{(2)}\}$ exclusively
obtained by Gosper's Hack algorithm.


\begin{thebibliography}{99}
\bibitem{LTE} \textquotedblleft LTE; Evolved universal terrestrial radio
access (E-UTRA); Physical channels and modulation,\textquotedblright\ 3GPP,
Sophia Antipolis Cedex, France, TS 36.211 V15.7.0, Oct. 2019.

\bibitem{5G} \textquotedblleft NR; Physical channels and
modulation,\textquotedblright\ 3GPP, Sophia Antipolis Cedex, France, TS
38.211 V16.2, Jul. 2020.

\bibitem{Wifi} \textquotedblleft Part 22; Cognitive wireless RAN medium
access control (MAC) and physical layer (PHY) specifications: Policies and
procedures for operation in the TV bands,\textquotedblright\ IEEE Standard
802.22-2011, Jul. 2011.

\bibitem{Sync 1} H. Minn, V. K. Bhargava, and K. K. B. Letaief,
\textquotedblleft A robust timing and frequency synchronization for OFDM
systems,\textquotedblright\ \emph{IEEE Trans. Wireless Commun.}, vol. 2, no.
4, pp. 822-839, Jul. 2003.

\bibitem{Sync 2} K. S. Kim, S. W. Kim, Y. S. Cho, and J. Y. Ahn,
\textquotedblleft Synchronization and cell-search technique using preamble
for OFDM cellular systems,\textquotedblright\ \emph{IEEE Trans. Veh. Technol.%
}, vol. 56, no. 6, pp. 3469-3485, Nov. 2007.

\bibitem{Sync 3} M. M. Gul, X. Ma, and S. Lee, \textquotedblleft Timing and
frequency synchronization for OFDM downlink transmissions using Zadoff-Chu
sequences,\textquotedblright\ \emph{IEEE Trans. Wireless Commun.}, vol. 14,
no. 3, pp. 1716-1729, Mar. 2015.

\bibitem{Sync 4} S. Johnson and O. A. Dobre, \textquotedblleft Time and
carrier frequency synchronization for coherent optical communication:
Implementation considerations, measurements, and
analysis,\textquotedblright\ \emph{IEEE Trans. Instrum. Meas.}, vol. 69, no.
8, pp. 5810-5820, Aug. 2020.

\bibitem{preamble} C.-D. Chung and W.-C. Chen, \textquotedblleft Preamble
sequence design for spectral compactness and initial synchronization in
OFDM,\textquotedblright\ \emph{IEEE Trans. Veh. Technol.}, vol. 67, no. 2,
pp. 1428-1443, Feb. 2018.

\bibitem{CA1} C.-D. Chung, W.-C. Chen, and C.-K. Yang, \textquotedblleft
Constant-amplitude sequences for spectrally compact OFDM training
waveforms,\textquotedblright\ \emph{IEEE Trans. Veh. Technol.}, vol. 69, no.
11, pp. 12974-12991, Nov. 2020.

\bibitem{Wei pilot} W.-C. Chen and C.-D. Chung, \textquotedblleft Spectrally
efficient OFDM pilot waveform for channel estimation,\textquotedblright\ 
\emph{IEEE Trans. Commun.}, vol. 65, no. 1, pp. 387-402, Jan. 2017.

\bibitem{pilot2} W.-C. Chen, C.-K. Yang, P.-T. Chi, and C.-D. Chung,
\textquotedblleft Pilot sequence design for spectral compactness and channel
estimation in OFDM,\textquotedblright\ in \emph{Proc. IEEE Veh. Technol.
Conf.}, Honolulu, track 7A, pp. 1-5, Sep. 2019.

\bibitem{Estimation 1} R. Negi and J. Cioffi, \textquotedblleft Pilot tone
selection for channel estimation in a mobile OFDM system,\textquotedblright\ 
\emph{IEEE Trans. Consumer Electron.}, vol. 44, no. 3, pp. 1122-1128, Aug.
1998.

\bibitem{Estimation 2} P. Stoica and O. Besson, \textquotedblleft Training
sequence design for frequency offset and frequency-selective channel
estimation,\textquotedblright\ \emph{IEEE Trans. Commun.}, vol. 51, no. 11,
pp. 1910-1917, Nov. 2003.

\bibitem{YL sequence} S. Yu and J.-W. Lee, \textquotedblleft Channel
sounding for multi-user massive MIMO in distributed antenna system
environment,\textquotedblright\ \emph{Electronics}, vol. 8, no. 1, pp. 1-14,
Jan. 2019.

\bibitem{PUCCH} L. Kundu, G. Xiong, and J. Cho, \textquotedblleft Physical
uplink control channel design for 5G new radio,\textquotedblright\ in \emph{%
Proc. IEEE 5G World Forum (5GWF)}, Silicon Valley, pp. 233-238, Nov. 2018.

\bibitem{SSB} J. Y. Han, O. Jo, and J. Kim, \textquotedblleft Exploitation
of channel-learning for enhancing 5G blind beam index
detection,\textquotedblright \emph{\ IEEE Trans. Veh. Technol.}, vol. 71,
no. 3, pp. 2925-2938, Mar. 2022.

\bibitem{RA 1} T. Kim, I. Bang, and D.-K. Sung, \textquotedblleft An
enhanced PRACH preamble detector for cellular IOT
communications,\textquotedblright\ \emph{IEEE Commun. Lett.}, vol. 21, no.
12, pp. 2678-2681, Dec. 2017.

\bibitem{RA 2} L. Zhen \emph{et al.}, \textquotedblleft Random access
preamble design and detection for mobile satellite communication
systems,\textquotedblright\ \emph{IEEE J. Sel. Areas Commun.}, vol. 36, no.
2, pp. 280-291, Feb. 2018.

\bibitem{RA 3} B. Liang, Z. He, K. Niu, B. Tian, and S. Sun,
\textquotedblleft The research on random access signal detection algorithm
in LTE systems,\textquotedblright\ in \emph{Proc. IEEE Int. Symp. Microwave,
Antenna, Propag. and EMC Technol. for Wireless Commun.}, Chengdu, China, pp.
115-118, Dec. 2013.

\bibitem{RA 4} A.-E. Mostafa, \emph{et al.}, \textquotedblleft Aggregate
preamble sequence design and detection for massive IOT with deep
learning,\textquotedblright\ \emph{IEEE Trans. Veh. Technol.}, vol. 70, no.
4, pp. 3800-3816, Apr. 2021.

\bibitem{RA 5} L. Zhen, H. Kong, Y. Zhang, W. Wang, and K. Yu,
\textquotedblleft Efficient collision detection based on Zadoff-Chu
sequences for satellite-enabled M2M random access,\textquotedblright\ in 
\emph{Proc. IEEE Int. Conf. Commun.}, Montreal, QC, Canada, pp. 1-6, Aug.
2021.

\bibitem{MIMO CS 1} S. Ali, Z. Chen, and F. Yin, \textquotedblleft Design of
orthogonal uplink pilot sequences for TDD massive MIMO under pilot
contamination,\textquotedblright\ \emph{J. Commun.}, vol. 12, no. 1, pp.
40-48, Jan. 2017.

\bibitem{MIMO CS 2} L. G. Giordano \emph{et al.,} \textquotedblleft Uplink
sounding reference signal coordination to combat pilot contamination in 5G
massive MIMO,\textquotedblright\ in \emph{Proc. IEEE Wireless Commun. Netw.
Conf.}, Barcelona, Spain, pp. 1-6, Apr. 2018.

\bibitem{MIMO CS 3 (LS)} F. Yang, P. Cai, H. Qian, and X. Luo,
\textquotedblleft Pilot contamination in massive MIMO induced by timing and
frequency errors,\textquotedblright\ \emph{IEEE Trans. Wireless Commun.},
vol. 17, no. 7, pp. 4477-4492, Jul. 2018.

\bibitem{Chu} D. C. Chu, \textquotedblleft Polyphase codes with good
periodic correlation properties,\textquotedblright\ \emph{IEEE Trans.
Inform. Theory}, vol. 18, pp. 531-532, Jul. 1972.

\bibitem{Popovic} B. M. Popovic, \textquotedblleft Generalized chirp-like
polyphase sequences with optimum correlation properties,\textquotedblright\ 
\emph{IEEE Trans. Inform. Theory}, vol. 38, pp. 1406-1409, Jul. 1992.

\bibitem{CAZAC} J. J. Benedetto and J. J. Donatelli, \textquotedblleft
Ambiguity function and frame-theoretic properties of periodic
zero-autocorrelation waveforms,\textquotedblright\ \emph{IEEE J. Select.
Topics Signal Process.}, vol. 1, no. 1, pp. 6-20, Jun. 2007.

\bibitem{PRACH Cell} R.-A. Pitaval, B. M. Popovi\'{c}, P. Wang, and F.
Berggren, \textquotedblleft Overcoming 5G PRACH capacity shortfall:
Supersets of Zadoff--Chu sequences with low-correlation
zone,\textquotedblright\ \emph{IEEE Trans. Commun.}, vol. 68, no. 9, pp.
5673-5688, Sep. 2020.

\bibitem{Adj channel 1} M. Faulkner, \textquotedblleft The effect of
filtering on the performance of OFDM systems,\textquotedblright\ \emph{IEEE
Trans. Veh. Technol.}, vol. 49, no. 5, pp. 1877-1884, Sep. 2000.

\bibitem{Adj channel 2} C.-D. Chung, \textquotedblleft Spectrally precoded
OFDM,\textquotedblright\ \emph{IEEE Trans. Commun.}, vol. 54, no. 12, pp.
2173-2185, Dec. 2006.

\bibitem{Adj channel 3} H.-M. Chen, W.-C. Chen, and C.-D. Chung,
\textquotedblleft Spectrally precoded OFDM and OFDMA with cyclic prefix and
unconstrained guard ratios,\textquotedblright\ \emph{IEEE Trans. Wireless
Commun.}, vol. 10, no. 5, pp. 1416-1427, May 2011.

\bibitem{Suppression 4} M. Ma, X. Huang, B. Jiao, and Y. J. Guo,
\textquotedblleft Optimal orthogonal precoding for power leakage suppression
in DFT-based systems,\textquotedblright\ \emph{IEEE Trans. Commun.}, vol.
59, no. 3, pp. 844-853, Mar. 2011.

\bibitem{Suppression 5} C.-D. Chung and K.-W. Chen, \textquotedblleft
Spectrally precoded OFDM without guard insertion,\textquotedblright\ \emph{%
IEEE Trans. Veh. Technol.}, vol. 66, no. 1, pp. 107-121, Jan. 2017.

\bibitem{Suppression 6} K. Hussain and R. Lopez-Valcarce, \textquotedblleft
Joint precoder and window design for OFDM sidelobe
suppression,\textquotedblright\ \emph{IEEE Commun. Lett.}, vol. 26, no. 12,
pp. 3044-3048, Dec. 2022.

\bibitem{Omega Pattern} G. E. Andrews, \emph{The Theory of Partitions.}
Cambridge: Cambridge University Press, 1998.

\bibitem{Omega Pattern Alg} N. M. Chase, \textquotedblleft Global structure
of integer partitions sequences,\textquotedblright\ \emph{Electron. J. Comb.}%
, vol. 1, pp. 1-25, Apr. 2004.

\bibitem{Gosper's Hack} D. E. Knuth, \emph{The Art of Computer Programming},
vol. 4. Upper Saddle River, NJ, USA: Addison-Wesley, 2005.

\bibitem{Gospe's Hack Alg} A. M. Foggia, \textquotedblleft Massively
parallel approaches to frustrated quantum magnets,\textquotedblright\ \emph{%
Master in High Performance Computing}, vol. 4, pp. 1-46, Jan. 2019.

\bibitem{Polygon} I. Pinelis, \textquotedblleft Cyclic polygons with given
edge lengths: Existence and uniqueness,\textquotedblright\ \emph{J. Geom.},
vol. 82, no. 1-2, pp. 156-171, Aug. 2005.

\bibitem{Proakis} J. G. Proakis and M. Salehi, \emph{Digital Communications,
5th ed.} New York: McGraw-Hill, 2008.

\bibitem{Coherence bandwidth} T. S. Rappaport, \emph{Wireless Communications}%
, Upper Saddle River, NJ: Prentice Hall, 2001.

\bibitem{5G Channel} \textquotedblleft NR; Study on channel model for
frequencies from 0.5 to 100 GHz,\textquotedblright\ 3GPP, Sophia Antipolis
Cedex, France, TS 38.901 V16.1, Nov. 2020.
\end{thebibliography}
\end{document}